\documentclass[apj]{emulateapj}
\usepackage{mathptmx}

\def\gtorder{\mathrel{\raise.3ex\hbox{$>$}\mkern-14mu
             \lower0.6ex\hbox{$\sim$}}}
\def\ltorder{\mathrel{\raise.3ex\hbox{$<$}\mkern-14mu
             \lower0.6ex\hbox{$\sim$}}}

\slugcomment{Submitted to ApJ}

\shorttitle{Supernova precursors}

\shortauthors{Ofek et al.}

\begin{document}

\title{Precursors prior to Type IIn supernova explosions are common: precursor rates, properties, and correlations}
\author{Eran O. Ofek\altaffilmark{1},
Mark Sullivan\altaffilmark{2},
Nir J. Shaviv\altaffilmark{3},
Aviram Steinbok\altaffilmark{1},
Iair Arcavi\altaffilmark{1},
Avishay Gal-Yam\altaffilmark{1},
David Tal\altaffilmark{1},
Shrinivas R. Kulkarni\altaffilmark{4},
Peter E. Nugent\altaffilmark{5,6},
Sagi Ben-Ami\altaffilmark{1},
Mansi M. Kasliwal\altaffilmark{7},
S. Bradley Cenko\altaffilmark{8},
Russ Laher\altaffilmark{9},
Jason Surace\altaffilmark{9},
Joshua S. Bloom\altaffilmark{6},
Alexei V. Filippenko\altaffilmark{6}, and
Ofer Yaron\altaffilmark{1}
}
\altaffiltext{1}{Benoziyo Center for Astrophysics, Weizmann Institute
  of Science, 76100 Rehovot, Israel}
\altaffiltext{2}{School of Physics and Astronomy, University of Southampton, Southampton SO17 1BJ, UK}
\altaffiltext{3}{Racah Institute of Physics, The Hebrew University, 91904 Jerusalem, Israel}
\altaffiltext{4}{Cahill Center for Astronomy and Astrophysics, California Institute of Technology, Pasadena, CA 91125, USA}
\altaffiltext{5}{Lawrence Berkeley National Laboratory, 1 Cyclotron Road, Berkeley, CA 94720, USA}
\altaffiltext{6}{Department of Astronomy, University of California, Berkeley, CA 94720-3411, USA}
\altaffiltext{7}{Observatories of the Carnegie Institution for Science, 813 Santa Barbara St., Pasadena, CA 91101, USA}
\altaffiltext{8}{Astrophysics Science Division, NASA/Goddard Space Flight Center, Mail Code 661, Greenbelt, MD, 20771, USA}
\altaffiltext{9}{Spitzer Science Center, California Institute of Technology, M/S 314-6, Pasadena, CA 91125, USA}

\begin{abstract}
There is a growing number of supernovae (SNe),
mainly of Type IIn, which present an
outburst prior to their presumably final explosion. These precursors may affect
the SN display, and are likely related to some poorly charted phenomena
in the final stages of stellar evolution. Here we present a sample of 16 nearby
Type IIn SNe for which we have Palomar Transient Factory (PTF) observations
obtained prior to the SN explosion. By coadding these images taken prior
to the explosion in time bins, we search for precursor events. We find five
Type IIn SNe that likely have at least one possible precursor event
(PTF\,10bjb; SN\,2010mc [PTF\,10tel];
PTF\,10weh; SN\,2011ht; PTF\,12cxj), three of which
are reported here for the first time. For each SN we calculate the
control time (i.e., the amount of time our survey was able to detect an event
brighter than a given luminosity). Based on this analysis we find that
precursor events among SNe~IIn are common: assuming a 
homogeneous population, at the one-sided 99\% confidence level, more than
$50\%$ of SNe~IIn have at least one pre-explosion outburst that is
brighter than $3\times10^{7}$\,L$_\odot$ (absolute magnitude $-14$)
taking place up to 1/3\,yr prior to the SN explosion. The average rate of such
precursor events during the year prior to the SN explosion is likely
larger than one per year
(i.e., multiple events per SN per year), and fainter precursors
are possibly even more common.
We also find possible correlations between the integrated luminosity
of the precursor, and the SN total radiated energy, peak luminosity,
and rise time. These correlations are expected if the precursors are
mass-ejection events, and the early-time light curve of these SNe is powered
by interaction of the SN shock and ejecta with optically thick
circumstellar material.

\end{abstract}

\keywords{
stars: mass-loss ---
supernovae: general ---
supernovae: individual: SN\,2010mc, PTF\,10bjb, SN\,2011ht, PTF\,10weh, PTF\,12cxj, SN\,2009ip}

\section{Introduction}
\label{sec:Introduction}

In flux-limited synoptic surveys, a few percent of all discovered
supernovae (SNe) show
narrow- to intermediate-width ($\sim 30$--$3000$\,km\,s$^{-1}$)
hydrogen and helium emission lines.
These are dubbed Type IIn SNe
(Schlegel 1990; Filippenko 1997; Kiewe et al. 2012)
and probably have massive progenitors (e.g., Gal-Yam \& Leonard 2009).
Their emission lines likely
originate from relatively abundant circumstellar material (CSM) around the
SN progenitor star (e.g., Chevalier \& Fransson 1994;
Chugai \& Danziger 1994; Chugai et al. 2003; Ofek et al. 2007, 2010, 2014a;
Smith et al. 2008, 2009),
ejected only a short time (of the order of months to decades) prior to
the SN explosion
(Dessart et al. 2009; Gal-Yam \& Leonard 2009; Ofek et al. 2010;
Ofek et al. 2013b).
Several theoretical mechanisms have been suggested to explain this
high mass loss in the final stages of stellar
evolution (e.g., Rakavy et al. 1967;
Woosley et al. 2007;
Arnett \& Meakin 2011;
Chevalier 2012;
Quataert \& Shiode 2012;
Shiode \& Quataert 2013;
Soker \& Kashi 2013).

Recently, five SNe with candidate pre-explosion outbursts
(precursors) have been detected a few months to years prior to the
SN explosion
(Foley et al. 2007; Pastorello et al. 2007;
Mauerhan et al. 2013a; Pastorello et al. 2013;
Ofek et al. 2013b;
Corsi et al. 2013;
Fraser et al. 2013).
In most cases these precursors were detected from SNe IIn,
or closely related events.

The frequency and properties of these precursors are critical for
pinpointing the eruption mechanisms and understanding their effect
on the eventual SN optical display, and may change our view of
the final stages of massive star evolution.  Therefore, we have
conducted a search for precursor events in a sample of nearby SNe~IIn
for which we have pre-explosion observations from the
Palomar Transient Factory
(PTF\footnote{http://ptf.caltech.edu/iptf/};
Law et al. 2009; Rau et al. 2009).
Our sample contains 16 SNe~IIn and we found
precursors events for five of the SNe in our sample.
For the first time, we estimate the rate of such events
and show that they are common. Furthermore, we
investigate possible correlations between the properties
of the precursors and the SNe.

The paper is organized as follows.
We describe the SN sample in \S\ref{sec:sample},
while the observations are presented in \S\ref{sec:Observations}.
The methodology of precursor selection is discussed
in \S\ref{sec:selection} and our
candidates are presented in \S\ref{sec:Indiv}.
We give our control-time estimate in \S\ref{sec:Control},
and the precursor rate in \S\ref{sec:rate}.
The CSM mass estimate is discussed in \S\ref{sec:CSM},
the correlations between the SNe and precursor
properties in \S\ref{sec:Corr}, and the results in \S\ref{sec:Disc}.

\section{Sample}
\label{sec:sample}

Our sample is based on SNe~IIn
that show intermediate-width
Balmer emission lines.
From inspection of many SNe spectra obtained by PTF,
we find that some SNe show this hallmark of SNe~IIn at early times (a 
few days after the explosion), but these lines disappear on a time scale
of a week.
It is possible that these SNe also suffer from a moderate mass-loss rate
prior to the explosion (e.g., Gal-Yam et al. 2014; Yaron et al. 2014).
However, 
we exclude from our sample objects for which the spectra evolve
into those of normal SNe~II a few weeks after
explosion.
Examples for such objects include PTF\,11iqb and PTF\,10uls (Ofek et al. 2013a).

Another important criterion for our SNe selection 
is that they have a large number of pre-explosion
images from the PTF survey
(Law et al.\ 2009; Rau et al.\ 2009).
The existence of a large number of images is critical,
as precursors may be relatively faint
and it is desirable to coadd images in order to get 
a limiting magnitude that is deeper than that of the nominal survey.
The SNe in our sample were found by the PTF
as well as amateur astronomers,
the Lick Observatory SN Search (LOSS; Li et al.\ 2000;
Filippenko et al. 2001),
and the Catalina Real-time Transient Survey (CRTS; Drake et al.\ 2009).
We selected only nearby SNe, found within 400\,Mpc, for which we
have a decent number of pre-explosion observations.
Table~\ref{tab:Samp} list the 16 SNe in our sample.
\begin{deluxetable*}{llllllllll}
\tablecolumns{10}
\tablewidth{0pt}
\tablecaption{SN sample}
\tablehead{
\colhead{Name}          &
\colhead{Type}          &
\colhead{$\alpha$(J2000)}      &
\colhead{$\delta$(J2000)}     &
\colhead{$M_{R,{\rm peak}}$}  &
\colhead{$z$}             &
\colhead{DM}            &
\colhead{$t_{{\rm rise}}$} &
\colhead{$t_{{\rm peak}}$} &
\colhead{FAP} \\
\colhead{}    &
\colhead{}    &
\colhead{(deg)} &
\colhead{(deg)} &
\colhead{(mag)} &
\colhead{}    &
\colhead{(mag)} &
\colhead{(day)} &
\colhead{(day)} &
\colhead{}
}
\startdata
SN\,2010jl & IIn     & 145.722236& $+$09.495037& $-$20.6  & 0.011 & 33.44 & 55474 & 55494  & 0.01 \\ 
SN\,2010jj & IIn     &  31.717743& $+$44.571558& $-$18.0: & 0.016 & 34.22 & 55512 & 55525  & 0.02 \\
PTF\,10achk            & IIn     &  46.489751& $-$10.522491& $-$18.7  & 0.0327& 35.77 & 55535 & 55550  & 0.00 \\
PTF\,10bjb             & IIn?     & 192.424667& $-$10.800159& $<-$16.4 & 0.026 & 35.27 & 55326 & 55434  & $...$ \\
SN\,2010bq & IIn     & 251.730659& $+$34.159641& $-$18.5  & 0.032 & 35.73 & 55296 & 55311  & 0.00 \\
PTF\,10gvf             & IIn     & 168.438496& $+$53.629126& $-$18.8  & 0.081 & 37.82 & 55321 & 55339  & 0.00 \\
SN\,2010mc & IIn     & 260.377817& $+$48.129834& $-$18.5  & 0.035 & 35.93 & 55427 & 55447  & 0.00 \\
PTF\,10weh             & IIn     & 261.710251& $+$58.852064& $-$20.7  & 0.138 & 39.06 & 55453 & 55504  & 0.02 \\
PTF\,11fzz             & IIn     & 167.694502& $+$54.105220& $-$20.3  & 0.082 & 37.85 & 55721 & 55783  & 0.00 \\
PTF\,12cxj             & IIn?     & 198.161181& $+$46.485090&  $-$17.3 & 0.036 & 35.96 & 56033 & 56048 & 0.01 \\
\hline
SN\,2011cc             & IIn     & 248.456000& $+$39.263528& $-$18.1: & 0.0319& 35.72 & 55624 & 55741 & 0.02 \\  
SN\,2011fx             & IIn     &   4.498167& $+$24.562778& $-$17.1: & 0.0193& 34.61 & 55780:& 55803: & 0.00 \\ 
SN\,2011ht             & IIn     & 152.044083& $+$51.849194& $-$16.8  & 0.004 & 30.96 & 55830:& 55880: & 0.00 \\ 
SN\,2011hw             & IIn/Ibn & 336.560583& $+$34.216417& $-$19.1: & 0.023 & 34.99 & 55860 & 55884: & $...$ \\ 
SN\,2011iw             & IIn     & 353.700833& $-$24.750444& $-$18.1: & 0.023 & 34.99 & 55819 & 55894: & 0.04 \\ 
SN\,2012as             & IIn     & 231.285500& $+$37.963722& $-$18.0: & 0.0297& 35.56 & 55968:& 55976: & 0.00   
\enddata
\tablecomments{The SNe sample. Type refer to SN type,
$\alpha$(J2000) and $\delta$(J2000) are the J2000.0 right ascension
and declination, respectively.
$M_{R, {\rm peak}}$ is the peak absolute $R$-band magnitude,
$z$ is the SN redshift,
DM is distance modulus,
$t_{{\rm rise}}$ is the MJD of the estimated start of the SN rise,
and $t_{{\rm peak}}$ is the MJD of the light curve peak.
Colon sign indicates an uncertain value.
The $t_{{\rm rise}}$ for SN\,2011iw is based on a detection in coadded PTF data
and it appears as a part of the fast rise following this detection.
FAP is the false-alarm probability to detect a precursor by coadding
images in 15-day bins as estimated using the bootstrap method
(see \S\ref{sec:tests}).
The values are based on 100 bootstrap simulations and therefore
truncated to two figures after the decimal point.
SNe with no data are those in which the precursor is clearly detected
in many individual images, and therefore, the bootstrap analysis
on the coadded data is ineffective (see \S\ref{sec:tests}).
SNe below the horizontal line were not detected by PTF,
although PTF have pre-explosion images of their locations.
\\
{\it References:}\\
SN\,2010jl: PTF\,10aaxf; Newton et al.\ (2010); Stoll et al.\ (2011);
Smith et al. (2011); Chandra et al. (2012); Ofek et al. (2013a; 2014a).\\
SN\,2010jj: PTF\,10aazn; Rich (2010); Silverman et al.\ (2010). \\
PTF\,10achk: reported here for the first time. \\
PTF\,10bjb: reported here for the first time. \\
SN\,2010bq: PTF\,10fjh; Duszanowicz (2010); Challis et al.\ (2010); Ofek et al. (2013a). \\
PTF\,10gvf: reported here for the first time. \\
PTF\,10tel: Ofek (2012); Ofek et al. (2013a; 2013b). \\
PTF\,10weh: reported here for the first time. \\
PTF\,11fzz: reported here for the first time. \\
PTF\,12cxj: reported here for the first time. \\
SN\,2011cc: Mason et al. (2011). \\
SN\,2011fx: Ciabattari et al. (2011). \\
SN\,2011ht: Boles et al.\ (2011); Prieto et al.\ (2011); Roming et al (2012); Mauerhan et al. (2013b)\\ 
SN\,2011hw: Dintinjana et al.\ (2011). \\ 
SN\,2011iw: Mahabal et al.\ (2011). \\
SN\,2012as: Jin et al. (2012). \\
}
\label{tab:Samp}
\end{deluxetable*}

\section{Observations}
\label{sec:Observations}

We used PTF observations of the SNe in our sample.
The PTF data reduction is described by Laher et al. (in prep.),
and the photometric calibration is discussed by Ofek et al. (2012a, 2012b).
Our search is based on image subtraction, and
the flux residuals in the individual image subtractions
for all the SNe in our sample are listed in Table~\ref{tab:MagFlux}.
\begin{deluxetable*}{lllllrlrl}
\tablecolumns{9}
\tablewidth{0pt}
\tablecaption{SN Observations}
\tablehead{
\colhead{Name}               &
\colhead{Band}               &
\colhead{MJD-$t_{{\rm rise}}$} &
\colhead{MJD}                &
\colhead{Mag}                &
\colhead{Mag Err}            &
\colhead{Lim Mag}            &
\colhead{Flux}               &
\colhead{Flux Err}           \\
\colhead{}    &
\colhead{}    &
\colhead{(day)} &
\colhead{(day)} &
\colhead{(mag)} &
\colhead{(mag)} &
\colhead{(mag)} &
\colhead{(counts)} &
\colhead{(counts)}
}
\startdata
PTF10aaxf &$R$&$   -232.84900$&  55241.15100 &  81.555&$   -0.913$&   21.441 &$      -66.4$&       55.8\\
PTF10aaxf &$R$&$   -232.49900$&  55241.50100 &  82.323&$   -0.985$&   20.590 &$     -134.7$&      122.2\\
PTF10aaxf &$R$&$   -229.61100$&  55244.38900 &  21.890&$    0.392$&   21.802 &$      110.7$&       40.0\\
PTF10aaxf &$R$&$   -229.56800$&  55244.43200 &  82.112&$   -0.653$&   21.247 &$     -110.9$&       66.7\\
PTF10aaxf &$R$&$   -223.68700$&  55250.31300 &  81.913&$   -1.045$&   20.936 &$      -92.3$&       88.8
\enddata
\tablecomments{Photometric measurements and flux residuals of the the SNe in our sample. Magnitude are calculated in laptitudes, and they have meaning
only when smaller than the limiting magnitude.
This table is published in its entirety in the electronic version of ApJ.
A portion of the full table is shown here for
guidance regarding its form and content.
}
\label{tab:MagFlux}
\end{deluxetable*}

\begin{figure*}
\centerline{\includegraphics[width=18cm]{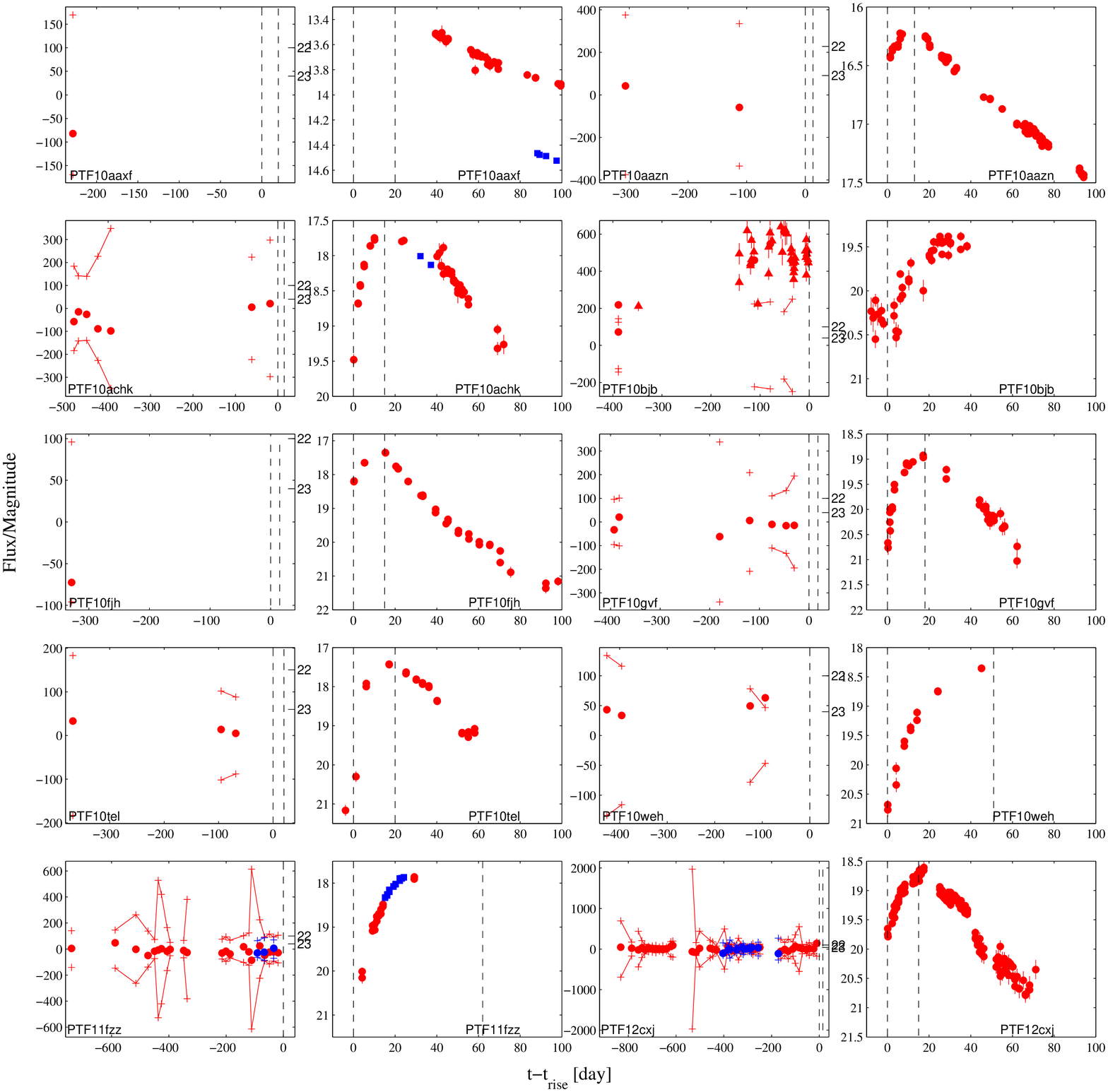}}
\caption{The light curves of ten SNe (names are listed in each panel).
For each SN two plots are shown (side by side in the first and second columns
and in the third and fourth columns).
Columns two and four show the light curve
after the SN explosion, while the panels in columns one and three
give only the coadded flux residual (relative to the reference image)
prior to the SN explosion (if there are more than
five observations per time bin).
The two vertical dashed lines show the assumed explosion date
($t_{{\rm rise}}$)
and rough maximum-luminosity date
($t_{{\rm peak}}$; listed in Table~\ref{tab:Samp}).
Time is measured relative to $t_{{\rm rise}}$.
The red circles represent $R$-band observations while the blue
squares show $g$-band data.
All of the measurements are in the PTF magnitude system
(Ofek et al. 2012a; 2012b).
The zero point of the flux residuals (first and third columns)
is 27, with the exception of
PTF\,10aazn (27.895),
PTF\,10bjb (27.14),
and PTF\,10tel (27.442).
The tick marks on the right-hand axes of the first and third columns
show the fluxes that correspond to PTF magnitudes of 22 and 23.
In the first and third columns, filled symbols show the
flux measurements, in 15-days bins.
Only bins containing six or more measurements are used.
The plus signs represent
the 5$\sigma$ upper and lower limits,
as estimated from the bootstrap technique in each bin (Efron 1982).
If plus signs are found in consecutive bins they are connected
by a solid line.
\label{fig:LCall1}}
\end{figure*}

\begin{figure*}
\centerline{\includegraphics[width=18cm]{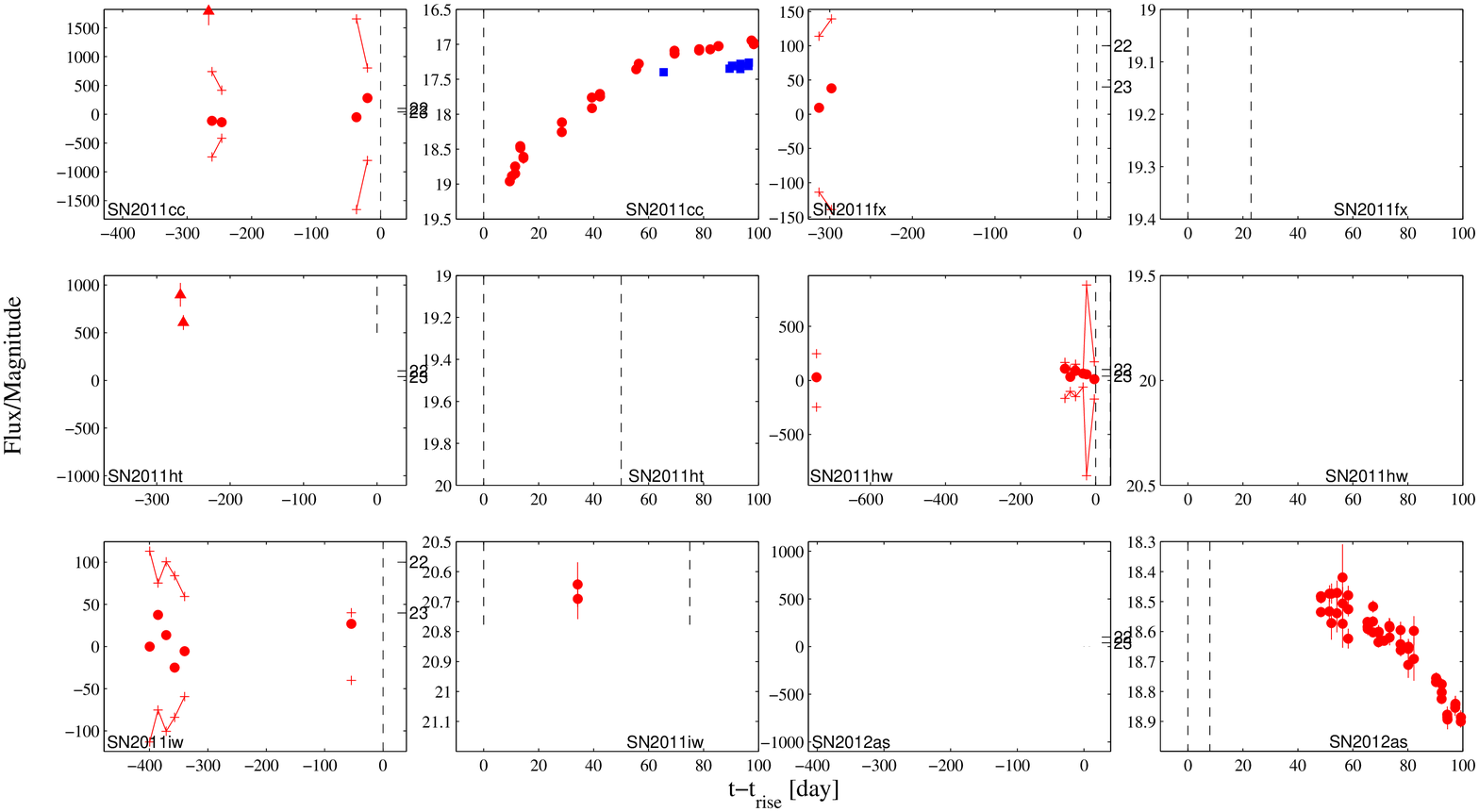}}
\caption{Like Figure~\ref{fig:LCall1} but
for six additional SNe.
\label{fig:LCall2}}
\end{figure*}

Figures~\ref{fig:LCall1}--\ref{fig:LCall2} show the light curves
before explosion (first and third columns) and after explosion (second
and fourth columns) of all the SNe in our sample.  The pre-explosion
light curve shows the median flux, relative to the reference image
flux, in 15-days bins.  Only bins containing $\ge6$ measurements are
presented.  The ``+'' signs shows the lower and upper 5$\sigma$ error
relative to the reference image, while the solid lines connect
consecutive bins.  The errors where calculated using the bootstrap
error on the mean.  The vertical dashed lines show the estimated
explosion time and time of maximum light (see Table~\ref{tab:Samp}).


We obtained spectra of our SNe
using various telescopes, and 
the log of selected observations is presented in
Table~\ref{tab:LogSpec}. Some of the spectra are presented in
this paper, while the rest are available electronically
from the WISeREP
website\footnote{http://www.weizmann.ac.il/astrophysics/wiserep/} (Yaron \& Gal-Yam 2012).
\begin{deluxetable}{lllll}
\tablecolumns{5}
\tablewidth{0pt}
\tablecaption{Log of spectroscopic observations}
\tablehead{
\colhead{SN name}       &
\colhead{MJD}           &
\colhead{Telescope}     &
\colhead{Instrument}    \\
\colhead{}              &
\colhead{(day)}           &
\colhead{}              &
\colhead{}              
}
\startdata
SN\,2010jl  & 55505   & Keck-I       & LRIS     \\
SN\,2010jj  & 55505   & Keck-I       & LRIS     \\
            & 55544   & Lick 3\,m      & Kast     \\
PTF\,10achk & 55547   & UH88         & SNIFS    \\
PTF\,10bjb  & 55262   & Keck-I       & LRIS     \\
SN\,2010bq  & 55300   & Gemini-N     & GMOS     \\
PTF\,10gvf  & 55322   & Keck-I       & LRIS     \\
SN\,2010mc  & 55434   & Gemini-N     & GMOS     \\
            & 55442   & Lick 3\,m      & Kast     \\
            & 55449   & KPNO 4\,m      & RC Spec  \\
            & 55455   & Lick 3\,m      & Kast     \\
PTF\,10weh  & 55479   & KPNO 4\,m      & RC Spec  \\
            & 55502   & Lick 3\,m      & Kast     \\
            & 55503   & KPNO 4\,m      & RC Spec  \\
PTF\,11fzz  & 55736   & Hale 5\,m      & DBSP     \\
            & 55873   & WHT          & ISIS     \\
PTF\,12cxj  & 56035   & Gemini-N     & GMOS     
\enddata
\tablecomments{MJD is the observation modified Julian day.
\label{tab:LogSpec}}
\end{deluxetable}

\section{Precursor Candidate Selection}
\label{sec:selection}

In this section we describe the methods we used
to find the precursor candidates.
In \S\ref{sec:Method} we present the search methods,
while in \S\ref{sec:tests} we discuss the
reliability of our methodology and the false-alarm 
probability of the precursor candidates.

\subsection{Detection Methods}
\label{sec:Method}

Our candidate precursor selection is based on two channels.
The first channel identifies precursors that were detected
in a single image prior to the SN explosion,
at the 6$\sigma$ level,
without the need to coadd data.
The second channel identifies precursors that were detected in the coadd data
at the 5$\sigma$ level, using 15-day bins and using only bins
that contain more than five flux measurements.

The second channel significantly increases the sensitivity of PTF to
precursor events by effectively coadding images of a SN
location in time bins of 15\,days.
These time bins often include
a large number of observations, extending to depths
beyond the nominal PTF survey limiting magnitude, and reaching 
an $R$-band limiting magnitude of $< 23.5$.
However, coadding images themselves in arbitrary time bins, and then
conducting image-subtraction analysis on each trial bin, is a
relatively expensive operation.  Instead, we carefully apply image
subtraction
to individual images, and for each image we save the flux residual
(negative or positive) at the location of the SN.
We can then coadd the scalar flux residuals in any temporal
combination we desire.
This method was used in the case of
SN\,2010mc (PTF\,10tel; Ofek et al. 2013b) and
PTF\,11qcj (Corsi et al. 2013).

In the first channel, the uncertainty (i.e., $\sigma$) was estimated based on
the Poisson noise propagated through the image-subtraction pipeline, while for
the second channel, we calculated errors in each bin using the
bootstrap technique (Efron 1982).  We note that in most cases the
bootstrap errors are consistent with the uncertainties derived based on the
standard deviation of the points in each bin, and the expected Poisson
noise.  Therefore, our bootstrap error estimate suggests that the
statistical uncertainties produced by the subtraction pipeline are
realistic.

In order to keep our search uniform we use only the PTF $R$-band data
for our search and analysis, as PTF was primarily an $R$-band search.
However, we also show the $g$-band data in the various plots where
available; the amount of $g$-band data is small in comparison to the
$R$-band.

Candidate precursor events detected via one of the channels
are discussed in \S\ref{sec:Indiv}.
In the initial search we do~not attempt to use different
time bins.
This decision was made in order to limit
the number of statistical experiments,
which may affect the significance of our results.
We note, however, that careful examination of specific events
with longer time bins may contain precursor events that are not
discussed here.
For example, in Corsi et al. (2013) we report on a possible
faint (absolute magnitude $-13$), several months long, brightening in the
light curve of the Type Ic PTF\,11qcj about 2.5\,yr prior to its explosion.

We note that several objects show points that are marginally
below the lower 5$\sigma$ error threshold.
It is possible that this is caused by real variability of the
progenitor (e.g., Szczygie{\l} et al. 2012), but here we concentrate on the outbursts
rather than possible dimmings (see also \S\ref{sec:PTF12cxj}).

\subsection{Tests}
\label{sec:tests}

\begin{figure}
\centerline{\includegraphics[width=7.5cm]{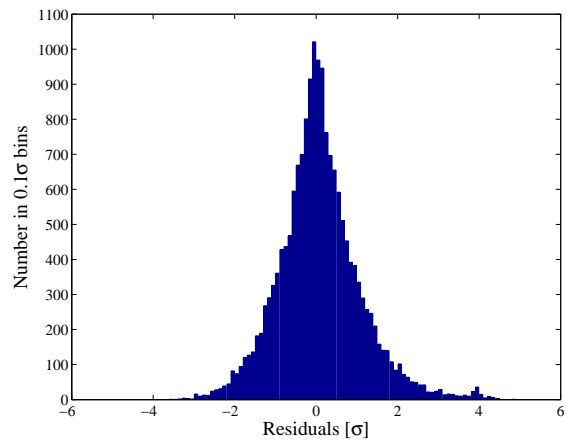}}
\caption{The distribution of flux residuals in $\sigma$ units,
at all the epochs and at 16 random positions around the position of PTF\,12cxj,
on the same host galaxy.
\label{fig:FakePositions_Res}}
\end{figure}

We performed several tests
to verify the reliability of our methodology,
especially against false alarms.
In order to test detections made by the first channel
(i.e., precursors detected in single images)
we extracted the
light curves at random positions
on top of the same host galaxy, but shifted in position
relative to the SN.
We found that typically the probability to get
a 6$\sigma$ detection is less than 0.1\% per image.
This is because
the noise is not distributed normally (e.g., there are
outliers) owing to occasional subtraction artefacts, cosmic rays, or
asteroids in the field of view.  For some SNe we have hundreds of
images, and therefore the probability for a detection is on the order
of a few percent per SN.
Therefore, we consider as good candidates
only events which have two consecutive detections (see \S\ref{sec:Indiv}).
Figure~\ref{fig:FakePositions_Res} illustrates the histogram
of ``random''-position flux residuals
in units of the flux-residual errors
(i.e., $\sigma$) for the case of random positions
around PTF\,12cxj.
It is apparent that even though there is a
small excess around positive residuals,
the probability of getting residuals which are bigger
than 6$\sigma$ is small.

In order to test detections made using the second channel
(coadding flux residuals in 15-day time bins),
for each SN we run 100 bootstrap simulations
in which we mixed the flux residuals and times,
and binned the data again in 15-day bins
with the same selection criteria.
The probability to have a single detection, based on these simulations,
is listed in column FAP (false-alarm probability) in Table~\ref{tab:Samp}.
This probability may approach several percent
for the entire sequence of images.
However, after passing the selection criteria,
each candidate is tested using various binning schemes, and
only sources that show two consecutive independent detections
are considered as good candidates (see \S\ref{sec:Indiv}).

We also search for correlations between the flux residuals and
airmass, and flux residuals and the amplitude of the expected
atmospheric refraction.  Indeed, we find marginally significant
correlations between these properties; however, they are too small to
affect our results.

To summarize,  tests of our methodology
suggest that the false-alarm probability for a
single-point precursor detection,
in an entire SN dataset, is $\sim5\%$ per object.
In order to avoid false detections, we consider as precursor
candidates only cases which show at least two detections clustered in
time (i.e., assuming the noise is not correlated).

\section{Candidate Precursor Events}
\label{sec:Indiv}

\begin{figure*}
\centerline{\includegraphics[width=18cm]{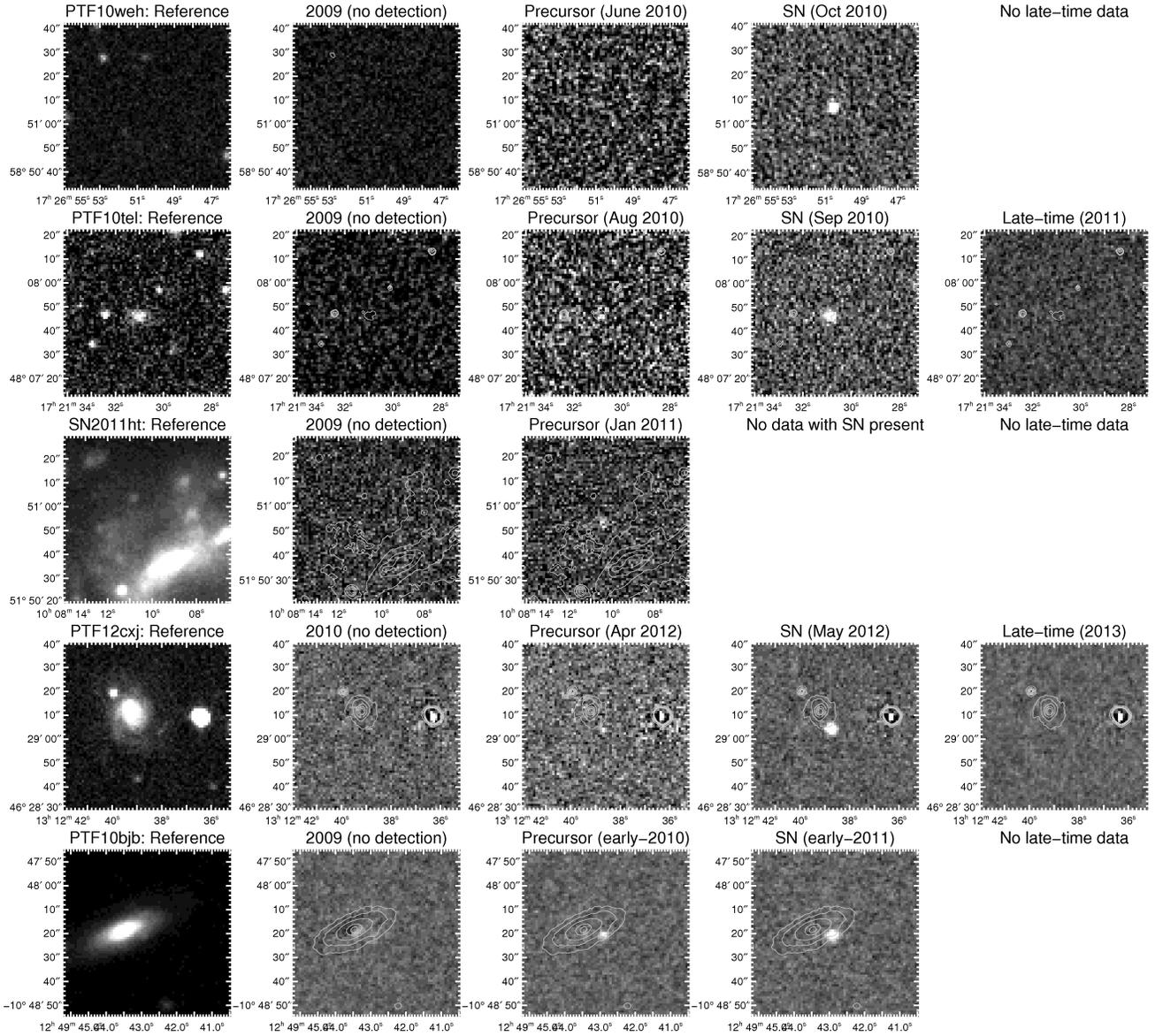}}
\caption{Images of the precursors, SN, and nondetections for the
SNe PTF\,10weh, PTF\,10tel, SN\,2011ht, PTF\,12cxj, and PTF\,10bjb
(from top to bottom). In each row, the five images present
(left to right) the reference image, the pre-SN explosion nondetection,
the brightest precursor for each object (where more than one was detected),
the SN, and a post-SN image (if available).
The contours from the reference image are also overplotted.
All of the precursors can be clearly seen, except PTF\,10weh
which is marginally detected.
A detailed inspection of the images of PTF\,10weh in which the
precursor is detected do not reveal any cosmetic problems
(e.g., cosmic rays) near the precursor location, and the balance of
evidence is that this precursor is real (see discussion
regarding false-alarm probability in \S\ref{sec:tests}).
However, in Figure~\ref{fig:Mcsm_corr} we also show the
correlations between the SN and precursor properties, excluding PTF\,10weh.
\label{fig:pre_image}}
\end{figure*}

Our search yielded several SNe with candidate precursors.
The precursors of PTF\,10bjb,
SN\,2010mc, and SN\,2011ht were clearly detected in individual images
(first channel), while PTF\,12cxj and PTF\,10weh are weakly detected
in coadded images (second channel).  Although these two weak events
pass all our tests and we consider them to be real, given the low
signal-to-noise ratio of the detections, we also discuss our results
assuming these two events are not real.
Images of all the precursors are presented in Figure~\ref{fig:pre_image}.

\subsection{PTF\,10tel}

PTF\,10tel is discussed in detail by Ofek et al. (2013b).  For
completeness, here we summarize the main properties of this event.  The
observations of PTF\,10tel show an outburst about 40\,days prior to
the probable explosion.  This precursor is detected via both the first
and second channels.  Our photometric and spectroscopic observations
suggest that this event is produced by an energetic outburst releasing
$\sim10^{-2}$\,M$_{\odot}$ at typical velocities\footnote{This is the
  larger $\sigma$ of a two-component Gaussian fit.}  of
2000\,km\,s$^{-1}$, and powered by at least $6\times10^{47}$\,erg of
energy.

\begin{figure}
\centerline{\includegraphics[width=7.5cm]{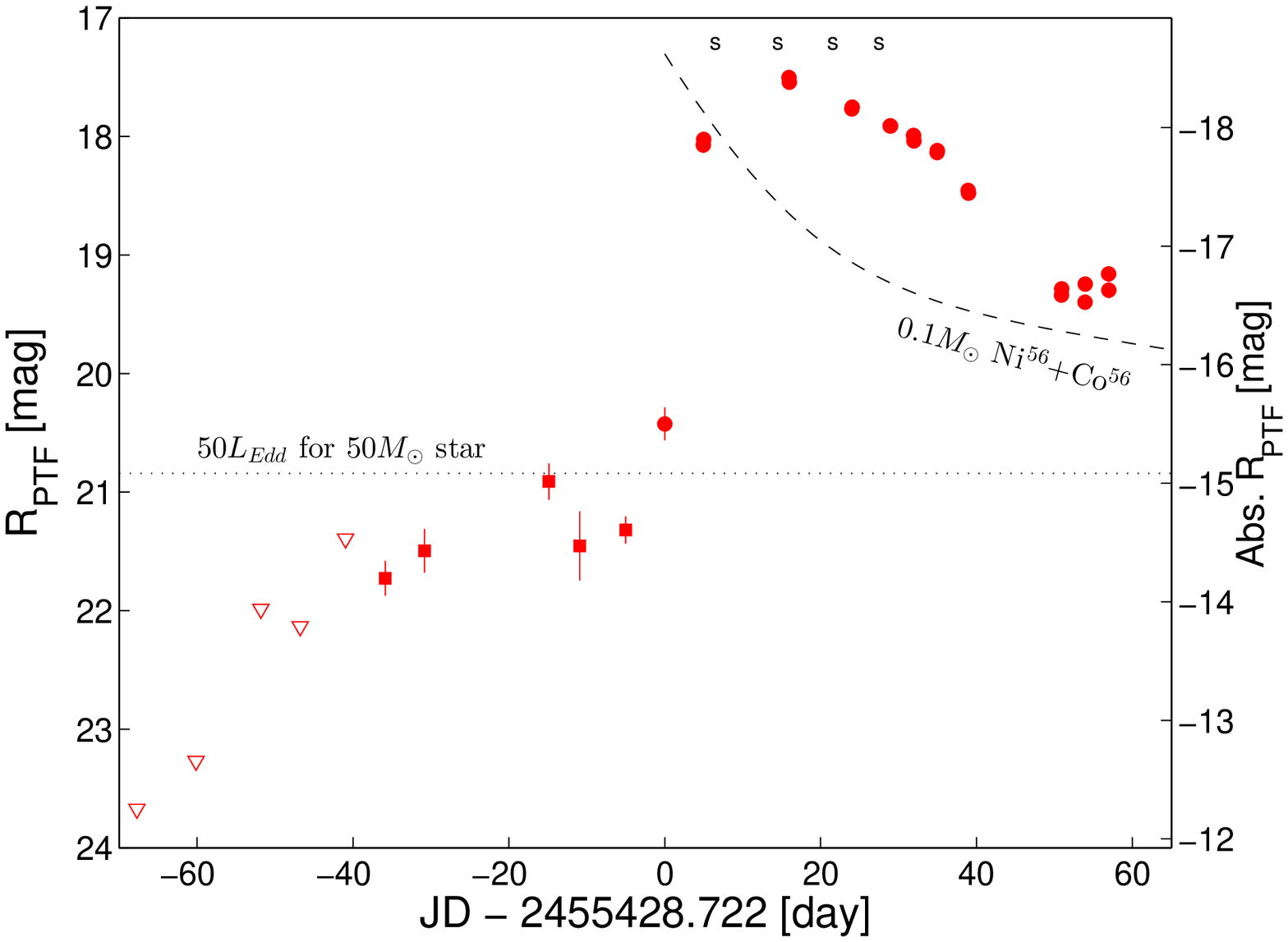}}
\caption{The light curve of SN\,2010mc (PTF\,10tel) as obtained with
  the Palomar 48-inch telescope.  The red circles are based on
  individual images, the squares are based on coadded images, while
  the empty triangles represent the 3$\sigma$ upper limits derived
  from coadded images.  The error bars represent the 1$\sigma$ errors.
  The object magnitudes are given in the PTF magnitude system.  The
  dashed line shows the expected luminosity from the radioactive decay
  of an ejected mass of 0.1\,M$_{\odot}$ of Ni$^{56}$, assuming that
  at late times the optical depth is sufficiently large to convert the
  radioactive energy to optical luminosity, but not so large that it
  goes into $P\,dV$ work.  This line represents an upper limit on the
  total amount of Ni$^{56}$ in the ejecta; it was set to coincide with
  the latest observation of the SN at JD $\approx$ 2,455,758.  The
  dotted line represents a bolometric luminosity equal to 50 times the
  Eddington luminosity for a 50\,M$_{\odot}$ star (an order of
  magnitude estimate of the mass of the progenitor assuming it is a
  massive star).  The right edges of the ``S'' symbols above the light
  curve indicate the epochs at which we obtained spectra (Figure
  adopted from Ofek et al. 2013b).
\label{fig:PTF10tel_LC}}
\end{figure}

For completeness we show in Figure~\ref{fig:PTF10tel_LC} the light curve
of PTF\,10tel from Ofek et al. (2013b).
The physical parameters of this SN and precursor
are listed in Table~\ref{tab:PrecProp}.

\begin{deluxetable*}{llllllllll}
\tablecolumns{10}
\tablewidth{0pt}
\tablecaption{Precursor Properties}
\tablehead{
\colhead{Name}             &
\colhead{$\Delta{t}$}      &
\colhead{$\delta{t}$}      &
\colhead{$L_{{\rm SN,peak}}$} &
\colhead{$E_{{\rm SN}}$}     &
\colhead{$L_{{\rm prec,peak}}$}&
\colhead{$E_{{\rm prec}}$}    &
\colhead{H$\alpha_{{\rm narrow}}$} &
\colhead{H$\alpha_{{\rm wide}}$}  &
\colhead{$\epsilon^{-1}M_{{\rm CSM}}$} \\
\colhead{}             &
\colhead{(day)}      &
\colhead{(day)}      &
\colhead{(erg\,s$^{-1}$)} &
\colhead{(erg)}     &
\colhead{(erg\,s$^{-1}$)} &
\colhead{(erg)}    &
\colhead{(km\,s$^{-1}$)} &
\colhead{(km\,s$^{-1}$)} &
\colhead{(M$_{\odot}$)}
}
\startdata
SN\,2010mc   & $-20$ & 30        & $6.9\times10^{42}$ & $2\times10^{49}$   &$2.6\times10^{41}$ & $6\times10^{47}$  & $\sim500$ & $\sim2000$ & 0.015 \\
PTF\,10bjb   & $-85$& 110       & $>8\times10^{41}$  & $>2.0\times10^{49}$&$3.3\times10^{41}$ & $2.4\times10^{48}$& $\sim300$    & $\sim600$     & 0.56    \\
SN\,2011ht   & $-205$& 210       & $1.5\times10^{42}$ & $2.5\times10^{49}$ &$1.5\times10^{40}$ & $2\times10^{47}$  & $260$        & $1800$        & 0.06  \\ 
PTF\,10weh   & $-80$& 40        & $5.3\times10^{43}$ & $7.2\times10^{50}$   &$1.9\times10^{42}$ & $5\times10^{48}$& $\sim100$    & $\sim1000$ & 0.5   \\
PTF\,12cxj-A & $-10$ & 7        & $2.3\times10^{42}$ & $8\times10^{48}$   &$2.2\times10^{41}$ &$9\times10^{46}$  & $\sim200$    & $\sim1200$ & 0.014 \\ 
PTF\,12cxj-B & $-700$& 15        &                   &                   & $4.3\times10^{40}$ &$6\times10^{46}$   &              &               &       \\ 
\hline
SN\,2009ip   & $-25$ & 50        & $4.6\times10^{42}$ & $3\times10^{49}$   &$2\times10^{41}$   &$8\times10^{47}$  & $\sim200$ & $\sim2000$ & 0.02\\
             & $-660$& $...$   &                   &                   &$1\times10^{41}$   & $...$         &              &               & $...$  \\
             & $-710$& 60        &                   &                  & $1\times10^{41}$   &$6\times10^{47}$  &              &               & 0.02     \\
             & $-1060$& 60       &                   &                  & $2\times10^{41}$   &$6\times10^{47}$  &              &               & 0.02     
\enddata
\tablecomments{Properties of the precursors and SNe.
H$\alpha_{{\rm narrow}}$ and H$\alpha_{{\rm wide}}$ are
the line width of the narrow and wide components of the
H$\alpha$ lines in units of the Gaussian $\sigma$ (multiply by 2.35 to get FWHM).
$\Delta{t}$ is the average time of the precursor.
All of the values are order-of-magnitude estimates.
The H$\alpha$ line widths for SN\,2011ht are adopted from Roming et al. (2012).
The values listed for SN\,2009ip are adopted from 
Foley et al. (2011), Smith et al. (2010), Drake et al. (2010),
Ofek et al. (2013c), and Margutti et al. (2014).
SNe listed above the horizontal line are in our sample
and used to estimate the precursor rates.
Events below the horizontal line are not in our sample,
but used in the correlation analysis (Figs.~\ref{fig:Mcsm_corr} and \ref{fig:Mcsmcont_corr}).
We note that SN\,2006jc (Pastorello et al. 2007;Foley et al. 2007)
was excluded from the correlation analysis since the
CSM in this SN is hydrogen deficient.
}
\label{tab:PrecProp}
\end{deluxetable*}

\subsection{SN\,2011ht}

SN\,2011ht was discovered by Boles (2011)
on 2011 Sep. 29.2 (UTC dates are used throughout this paper), 
at apparent magnitude 17.0.
Based on a spectrum obtained on 2011 Sep. 30,
Pastorello et al. (2011) suggested that it is a SN impostor sharing some
similarities with the eruption of the luminous blue variable UGC\,2773-OT
(Smith et al. 2010).  The spectrum shows narrow lines
of H, Ca\,II (H\&K and the near-infrared triplet, with P-Cyg profiles),
and a forest of narrow Fe\,II absorption features. Also, prominent Na\,I~D,
Sc\,II, and Ba\,II features are detected in absorption.
Prieto et al. (2011) obtained a further spectrum on 2011 Nov. 11.5.
They reported substantial evolution with respect to the initial
classification, with strong Balmer and weaker He\,I and Fe\,II 
emission lines superposed on a blue continuum.  Based on the spectrum and
the SN absolute magnitude, they suggested that it is likely a
SN~IIn.  This is supported by Roming et al. (2012), who reported on the
{\it Swift}-UVOT observations of this SN, detecting a 7\,mag
rise in the $UVW2$ band over 40\,days, peaking at a $u$-band magnitude
of about $-18$.

Fraser et al. (2013) reported on
an outburst peaking at a $z$-band absolute magnitude
of $-11.8$ detected $\sim 9$\,months prior to the explosion,
and with the last detection $\sim 4$ months before the explosion.
The event was detected by both PanSTARRS-1 and the Catalina Sky Survey.
The duration of the outburst is not well constrained, and it can be either
a single event that lasted for more than 4 months or multiple shorter events.

PTF observations detected the outburst in individual images,
suggesting that the outburst was already active 11\,months
prior to the explosion.
All of the PTF photometric
measurements are listed in Table~\ref{tab:MagFlux}.
We note that PTF did~not observe the SN itself.
Assuming there was a single precursor, the properties and energetics of this event
are listed in Table~\ref{tab:PrecProp}.

\subsection{PTF\,10bjb}

\begin{figure}
\centerline{\includegraphics[width=7.5cm]{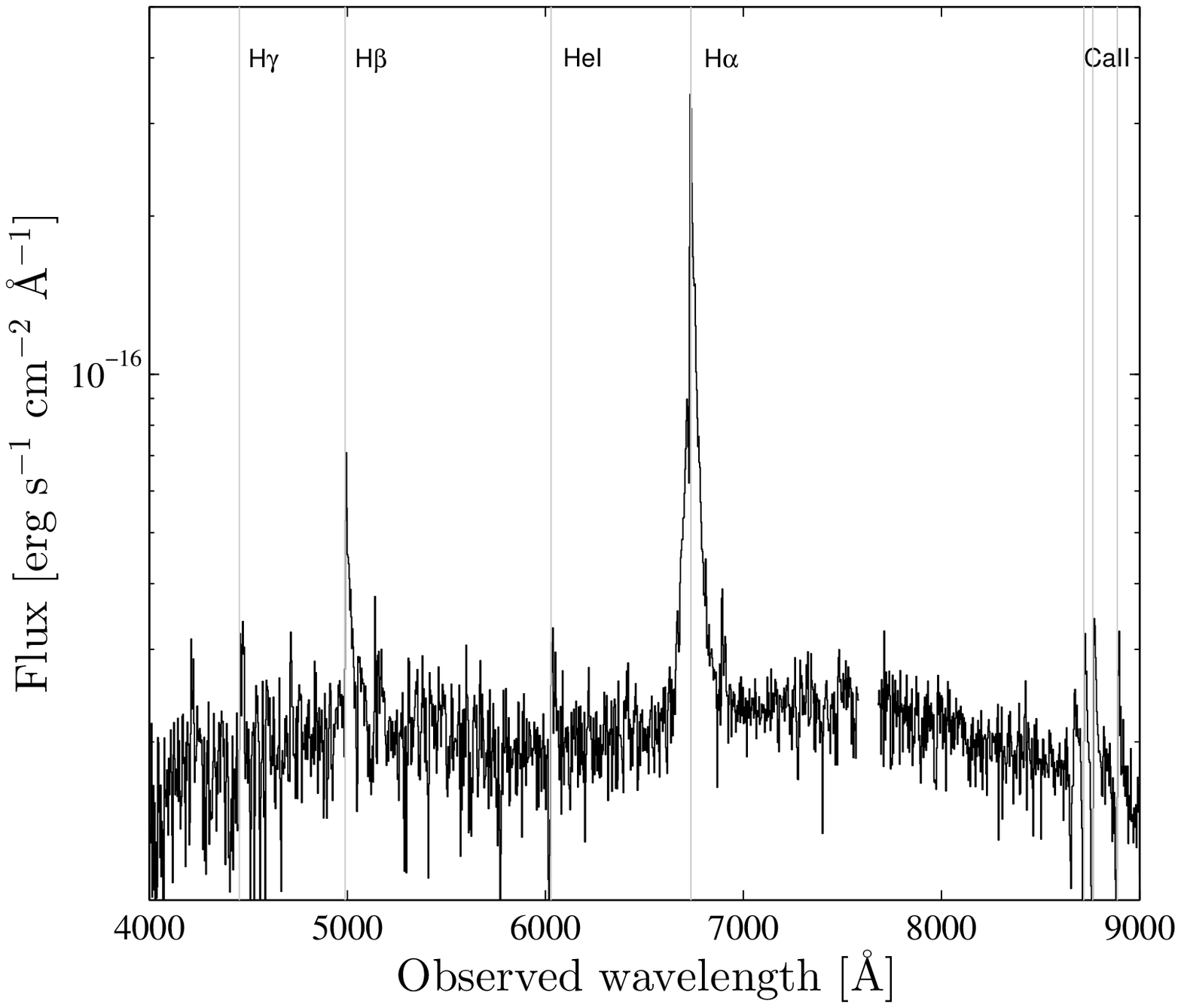}}
\caption{The spectrum of PTF\,10bjb as obtained by Keck-I/LRIS
on 2010 Mar. 7.
\label{fig:PTF10bjb_spec}}
\end{figure}

PTF\,10bjb was discovered by PTF on 2010 Feb. 16, and the only spectrum
obtained on 2010 Mar. 7 resembles those of luminous blue variables
and SNe~IIn.
Close inspection of the light curve suggests that the SN was discovered
while in a pre-explosion high state, and that the spectrum was obtained
prior to the explosion.
The spectrum shows Balmer emission lines, 
the broader component of which
has a velocity ($\sigma$) of about 600\,km\,s$^{-1}$.
Also detected are emission lines of He\,I and Ca\,II.
The He\,I $\lambda$5876 line shows a narrow P-Cygni profile with a velocity
of $\sim 600$\,km\,s$^{-1}$, while the H$\alpha$ line shows
a P-Cygni profile with a velocity of $\sim 300$\,km\,s$^{-1}$.
The spectrum of this event is presented in Figure~\ref{fig:PTF10bjb_spec}.

\begin{figure}
\centerline{\includegraphics[width=7.5cm]{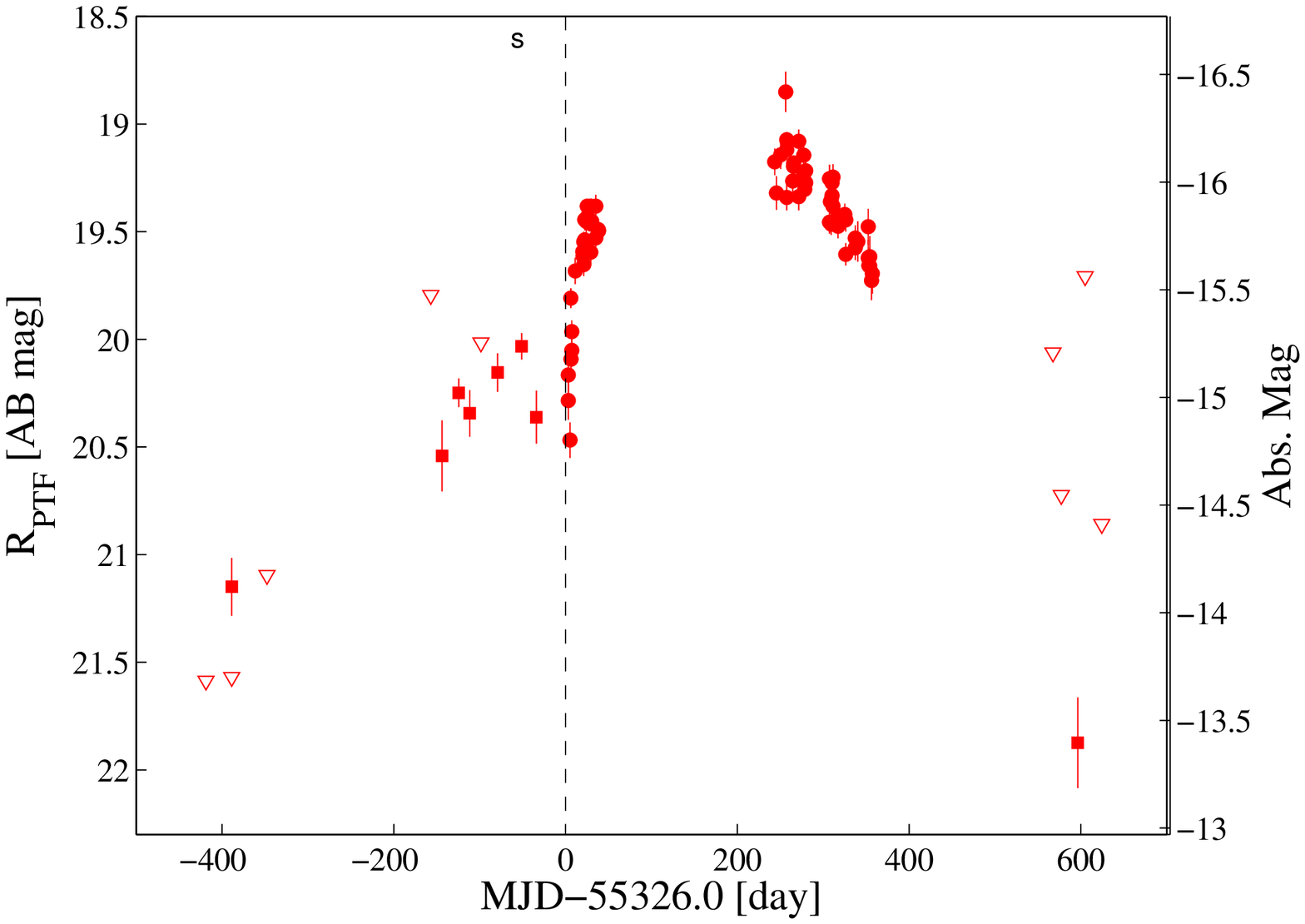}}
\caption{The light curve of PTF\,10bjb exhibits
a clear outburst lasting for $\sim 4$\,months
prior to the explosion (identified by the fast rise in
the light curve).
An additional possible detection is seen $\sim 400$\,days prior to
the explosion, but we do~not consider it to be a good candidate
since it is based on a single binned detection (see the discussion
about false-alarm probability in \S\ref{sec:tests}).
The upper limits corresponds to 5$\sigma$ bounds
calculated in 15-day bins.
\label{fig:PTF10bjb_LC_all}}
\end{figure}

The light curve of PTF\,10bjb is
shown in Figure~\ref{fig:PTF10bjb_LC_all}.
Between $\sim 80$\,days up to 0\,days prior to the assumed explosion time
of this SN, which is based on the fast rise in the SN light curve,
we detect a significant excess in the flux residuals.
This excess had a peak absolute magnitude of $-15.1$
and it lasted for $\sim 110$\,days.
Two years after $t_{{\rm rise}}$, the SN is detected at a flux level which is
a factor of $\sim 5$ dimmer than the pre-explosion outburst.
Assuming zero bolometric correction,
the peak bolometric luminosity of this outburst is
$\sim 3\times10^{41}$\,erg\,s$^{-1}$,
and its radiated energy is $\sim 2.6\times10^{48}$\,erg.

In addition, there is a possible detection of the progenitor
at an absolute magnitude of roughly $-14$ about a year prior to
the SN explosion.
However, given that there is a single detection at this time,
we do~not consider this to be a good precursor candidate (see \S\ref{sec:tests}).
The physical parameters of the SN and the precursor are listed in
Table~\ref{tab:PrecProp}.

\subsection{SN\,2011hw}

SN\,2011hw was discovered by Dintinjana et al. (2011)
on 2011 Nov. 18.7, at apparent magnitude 15.7.
Valenti et al. (2011) reported that a spectrum
obtained on 2011 Nov. 19.8
shows it to be similar to the transitional Type IIn/Ibn
SN\,2005la (Pastorello et al. 2008).  The spectrum
is blue and exhibits emission lines of H and He\,I.
The most prominent He\,I lines compete in strength with
H$\alpha$.  The FWHM velocity of H$\alpha$
is 2700\,km\,s$^{-1}$, while that of He\,I $\lambda$5876
is $\sim 2000$\,km\,s$^{-1}$.
We note that the apparent magnitude of this SN reported
by Dintinjana \& Mikuz (2011) corresponds to
an absolute magnitude of about $-19.3$.
Our initial search suggested that there is a detection
of an outburst $\sim 3$\,months prior to the SN explosion.
However, since this is based on a single detection,
we do~not consider this to be a good candidate.

\subsection{PTF\,10weh}

\begin{figure}
\centerline{\includegraphics[width=7.5cm]{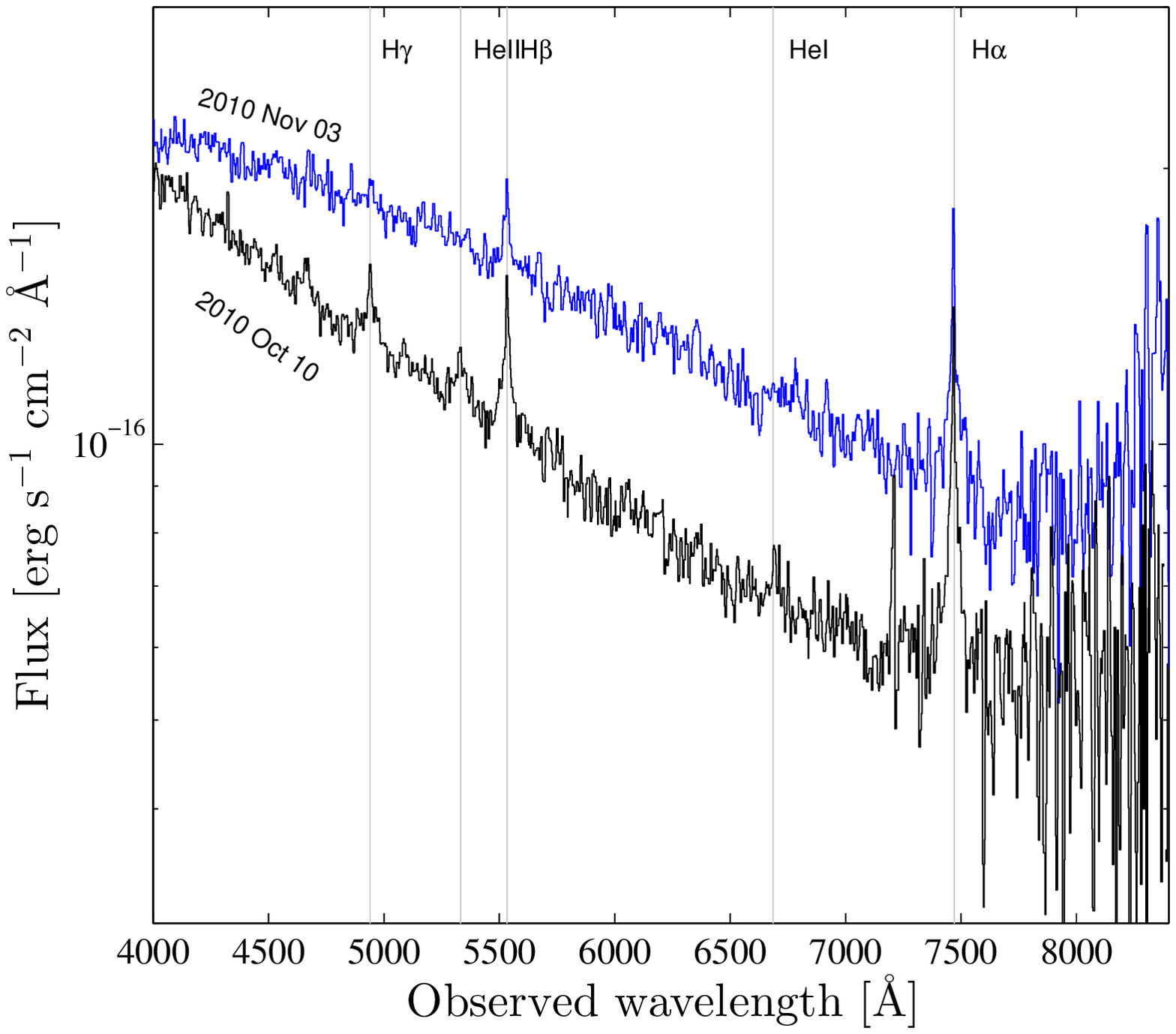}}
\caption{Spectra of PTF\,10weh as obtained using the KPNO 4\,m telescope
equipped with the RC spectrograph.
\label{fig:PTF10weh_spec}}
\end{figure}

PTF\,10weh was discovered by PTF on 2010 Sep. 22.14.
The SN brightened to a peak absolute magnitude of about $-20.7$
over $\sim 40$\,days.
A spectrum taken on 2010 Oct. 10 exhibits
a blue continuum together with Balmer emission lines as well
as He\,I and He\,II.
The widest component of the H$\alpha$ line has a velocity
($\sigma$) of $\sim1000$\,km\,s$^{-1}$.
The first two spectra of this source are shown in
Figure~\ref{fig:PTF10weh_spec}.

\begin{figure}
\centerline{\includegraphics[width=7.5cm]{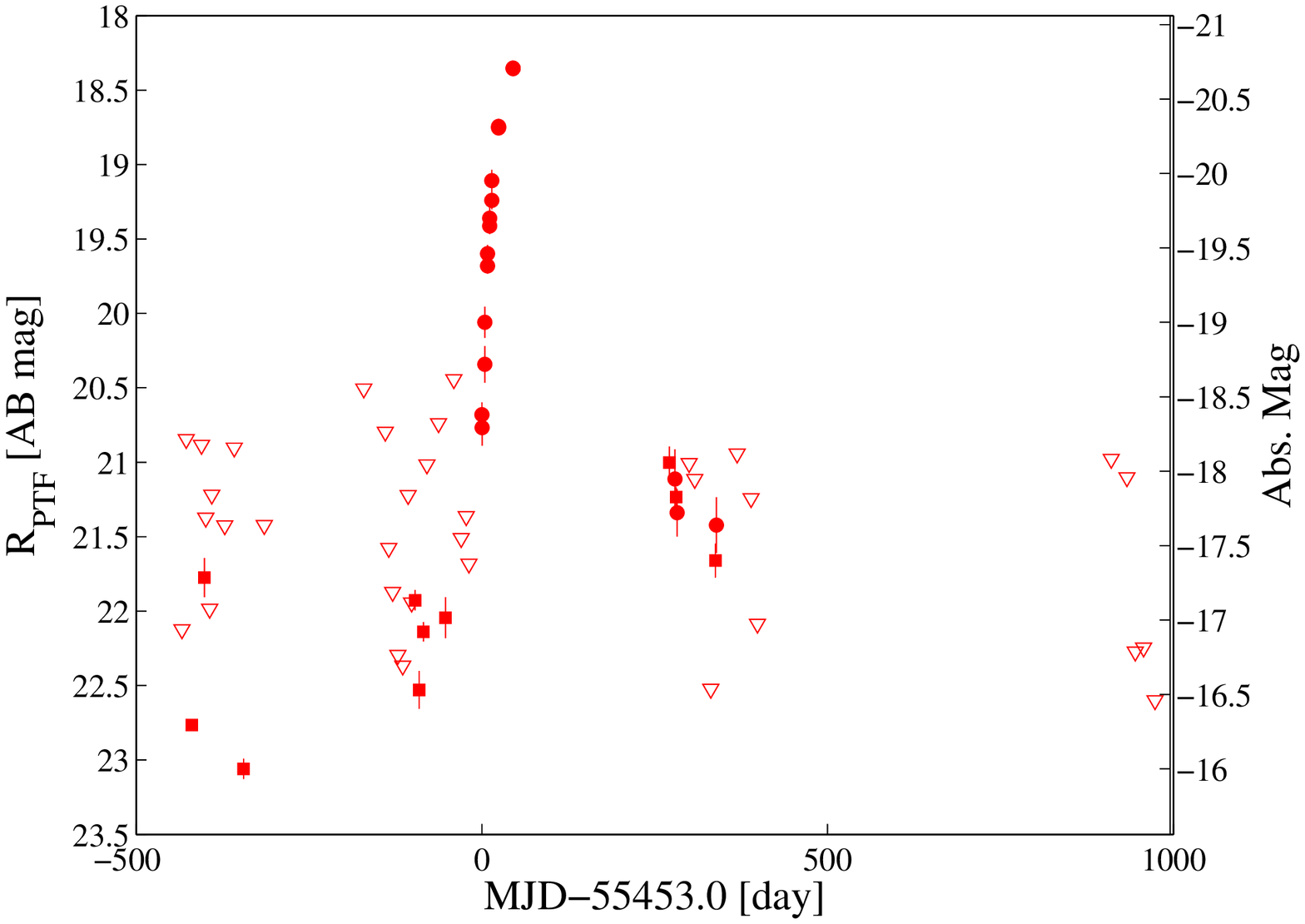}}
\caption{The PTF $R$-band light curve of PTF\,10weh.
Filled circles show individual measurements, while filled boxes
mark 3-day binned measurements with at least two measurements per bin.
The empty triangles denote 5$\sigma$ upper limits.
The bin size is 3\,days prior to the SN
explosion and 15\,days after the SN explosion.
\label{fig:PTF10weh_LC_all}}
\end{figure}

\begin{figure}
\centerline{\includegraphics[width=7.5cm]{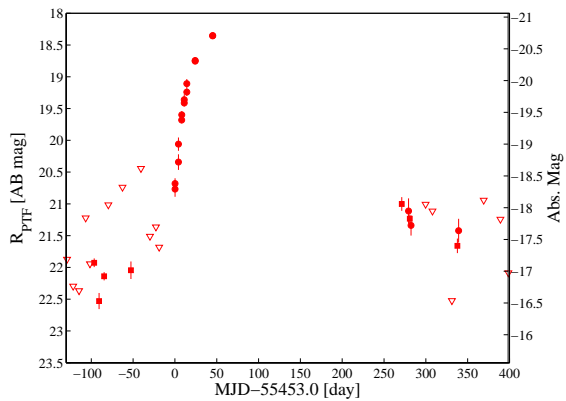}}
\caption{Close-up view of the PTF $R$-band light curve of PTF\,10weh around
the precursor candidate time (see also Fig.~\ref{fig:PTF10weh_LC_all}).
Filled circles denote individual measurements, while filled boxes
mark 3-day binned measurements with at least two measurements per bin.
The empty triangles are 5$\sigma$ upper limits.
The bin size is 3\,days prior to the SN
explosion and 15\,days after the SN explosion.
\label{fig:PTF10weh_LC_all_zoom}}
\end{figure}

\begin{figure}
\centerline{\includegraphics[width=7.5cm]{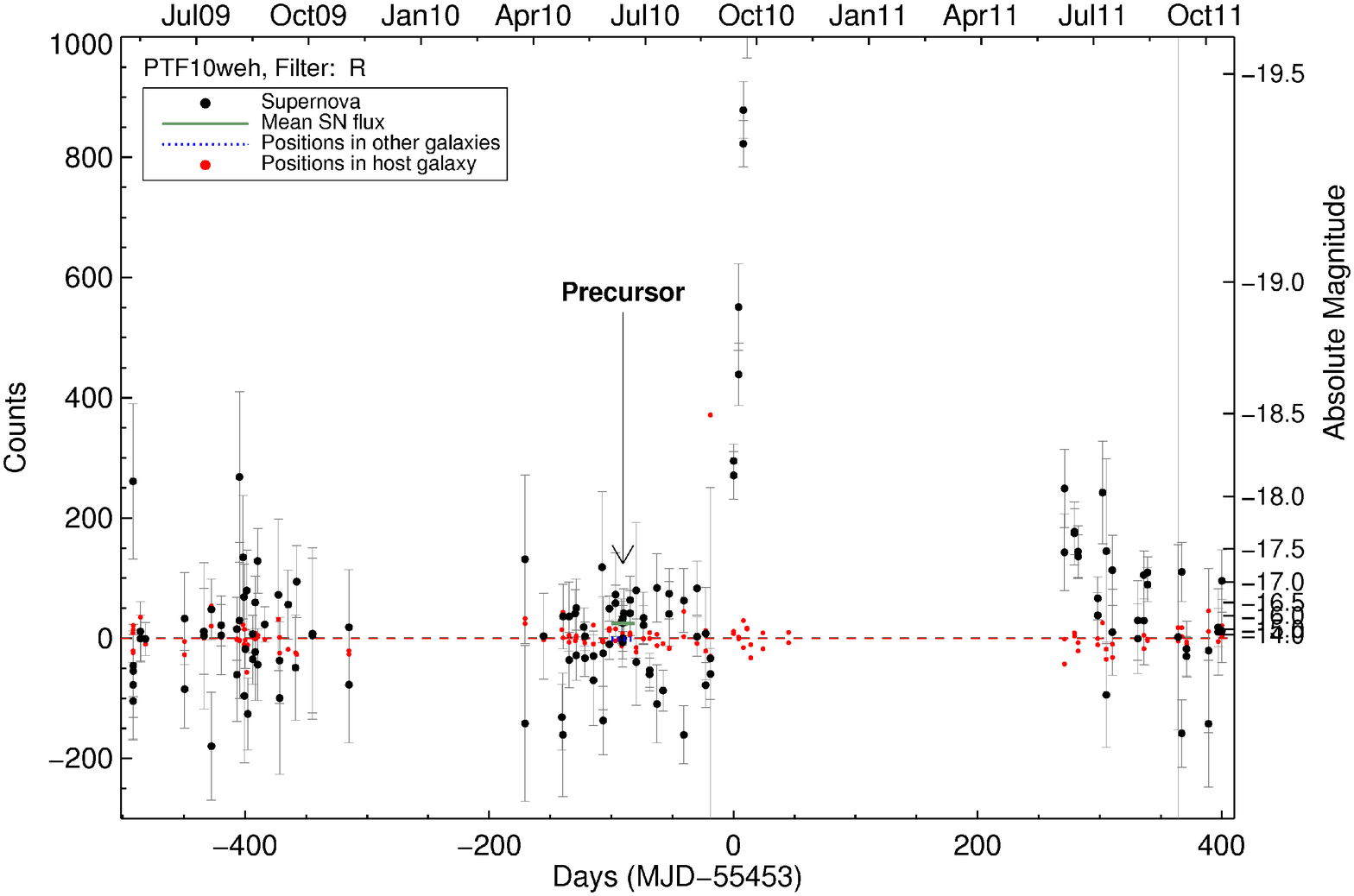}}
\caption{The counts-space light curve of SN PTF\,10weh.
Each black circle represents a point-spread function (PSF)
flux measurement on an individual PTF exposure after image subtraction.
The precursor is labelled, and the mean precursor flux is shown as a
green horizontal line, with the hashed area representing the uncertainty.
The blue line shows the average of random measurements in other
galaxies in the image, illustrating the typical flux uncertainty
and residual in the flux measurements after image subtraction.
The red circles show the average of random measurements in the
same host as the SN (where the host is extended enough to allow this test),
tracking the random subtraction noise that may affect the flux measurements.
The horizontal red line denotes the mean of the red circles, and is
consistent with zero flux as expected after image subtraction.
The absolute magnitude scale is shown on the right-hand axis for
reference, and the abscissa shows the time in days relative
to the SN $t_{{\rm rise}}$.
\label{fig:PTF10weh_diffstats_R}}
\end{figure}

The light curve of this SN and the candidate precursor event is
shown in Figure~\ref{fig:PTF10weh_LC_all}.
There is a possible precursor
$\sim 3$\,months prior to the SN rise, lasting
for $\sim40$\,days.
A close-up view of the precursor light curve is presented in
Figure~\ref{fig:PTF10weh_LC_all_zoom}.
Figure~\ref{fig:PTF10weh_diffstats_R} gives another version
of the precursor light curve, showing all of its individual
flux measurements,
along with flux measurements at random positions in
the host galaxy and other nearby galaxies.
Inspection of the data reveals that
the detection depends on the binning scheme.
However, given that there are four temporally adjacent
points that deviate by more than 5$\sigma$, we consider
this to be a good precursor candidate.
The absolute magnitude of this precursor is
about $-17$.
Assuming zero bolometric correction,
this corresponds to a luminosity
of $\sim 1.9\times10^{42}$\,erg\,s$^{-1}$.
The lower limit on the radiated energy of the precursor
is $\sim 5\times10^{48}$\,erg.
The physical parameters of the SN and its outburst are listed in
Table~\ref{tab:PrecProp}.

In addition, the pre-explosion light curve of this SN reveals
additional possible detections 11--14\,months prior
to the SN explosion.
However, given that these detections are not consecutive in time,
we do~not regard them as good precursor candidates.

\subsection{PTF\,12cxj}
\label{sec:PTF12cxj}

\begin{figure}
\centerline{\includegraphics[width=7.5cm]{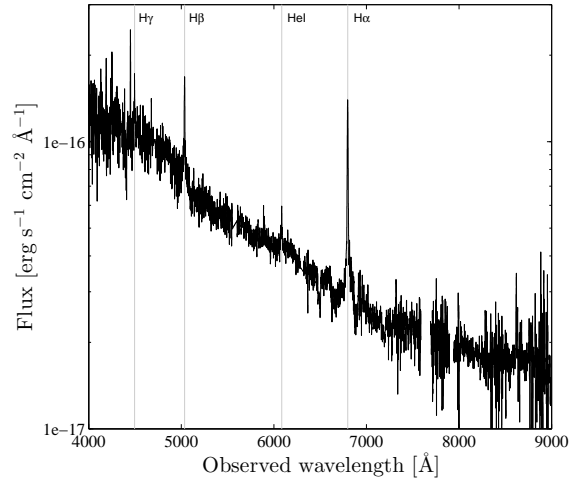}}
\caption{Spectrum of PTF\,12cxj as obtained with the Gemini-N telescope
equipped with the GMOS spectrograph.
\label{fig:PTF12cxj_spec}}
\end{figure}

PTF\,12cxj was discovered by PTF on 2012 Apr. 16.14.
The SN rose to an $R$-band absolute magnitude of $-17.2$ over
two weeks.
The spectrum of the SN, presented in Figure~\ref{fig:PTF12cxj_spec},
was obtained on 2012 Apr. 18 during the early rise of the SN.
The spectrum has Balmer and He\,I lines in emission.
Fitting a two Gaussian model to the H$\alpha$ line shows that,
the widest component has a velocity width ($\sigma$)
of 1200\,km\,s$^{-1}$.

\begin{figure}
\centerline{\includegraphics[width=7.5cm]{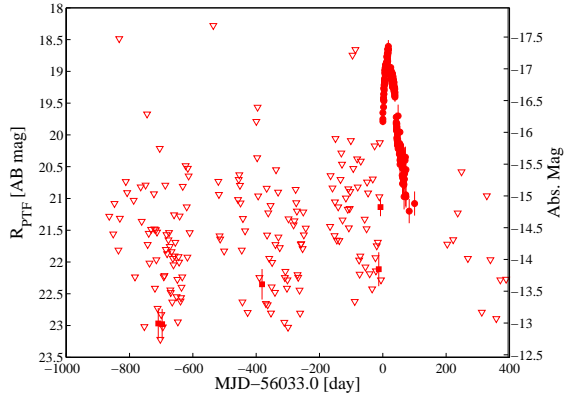}}
\caption{The PTF $R$-band light curve of PTF\,12cxj.
Filled circles denote individual measurements, while filled boxes
mark binned measurements.
The empty triangles show 4$\sigma$ upper limits.
The bin size is 2\,days prior to the SN explosion.
The minimum number of measurements in each bin is 4.
\label{fig:PTF12cxj_LC_all}}
\end{figure}

\begin{figure}
\centerline{\includegraphics[width=7.5cm]{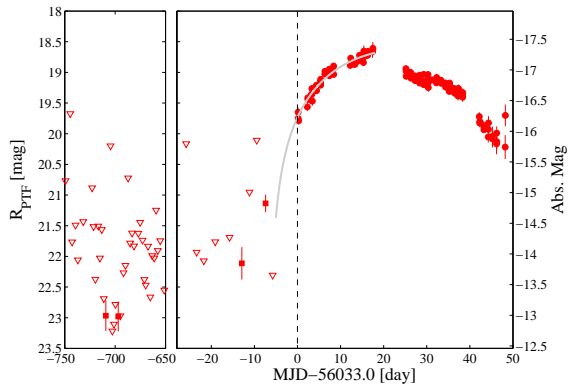}}
\caption{The light curve of PTF\,12cxj shows
two possible precursors (based on multiple consecutive
detections, where single detections are ignored).
The upper limits corresponds to 4$\sigma$ bounds
calculated in 2\,day bins.
Extrapolation of the light curve, based on a $t^{2}$ law (gray line),
suggests that the detections visible at around $-10$ days
took place before the SN explosion (see also Figure~\ref{fig:pre_image}).
Furthermore, there is a deep nondetection between these
detections and the assumed time of the SN explosion.
We note that even if we remove this event from our sample,
the rate of precursors does~not changed
by much, and the correlations (Fig.~\ref{fig:Mcsm_corr}) are still detected.
\label{fig:PTF12cxj_LC_all1}}
\end{figure}

This SN has a large number of pre-explosion observations,
and it shows a single detection $\sim 384$ days prior to its $t_{{\rm rise}}$.
Closer inspection reveals that this detection depends on the binning scheme,
and that with shorter bins of four days, multiple 
detections at an absolute magnitude of about $-13.5$ are seen.
Moreover, inspection of the $g$-band data shows a marginal detection
around $-435$\,days prior to $t_{{\rm rise}}$.
The full $R$-band light curve is illustrated in
Figure~\ref{fig:PTF12cxj_LC_all}.
The most convincing precursor candidate events are detected
about two weeks and 700\,days prior to $t_{{\rm rise}}$.
Given the large number of observations of this SN field,
the light curve in this plot utilized 2-day bins and
4$\sigma$ upper limits.
A close-up view of the light curve of the two precursors
is shown in Figure~\ref{fig:PTF12cxj_LC_all1}.
The $R$-band peak absolute magnitudes of these candidate
precursors are $-14.8$ and $-13$.
However, the candidate precursor two weeks prior to the explosion
is very close in time to the SN fast rise
and therefore may be regarded
as part of the SN rise.
As seen in Figure~\ref{fig:PTF12cxj_LC_all1},
the two detections are followed by
a nondetection which is about a magnitude deeper
than the possible detection at $-8$\,days prior to $t_{{\rm rise}}$.
Furthermore,
a $t^{2}$ extrapolation to the SN
rise flux suggests that the SN started
its rise after this candidate precursor.
We cannot rule out the possibility that this precursor
is part of the SN light curve (e.g., similar to SN\,2011dh; Arcavi et al. 2011).

\begin{figure}
\centerline{\includegraphics[width=7.5cm]{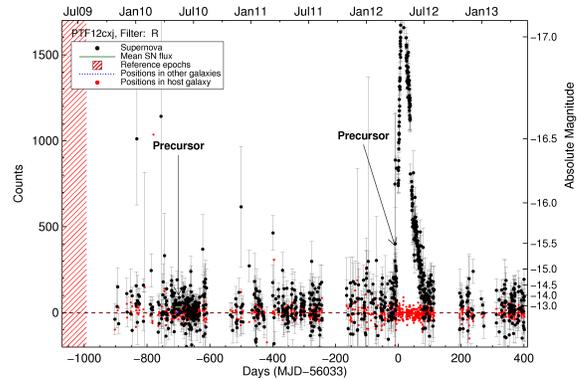}}
\caption{Like Figure~\ref{fig:PTF10weh_diffstats_R}, but for PTF\,12cxj.
\label{fig:PTF12cxj_diffstats_R}}
\end{figure}

As before, Figure~\ref{fig:PTF12cxj_diffstats_R} shows another version
of the precursor light curve, with all of the individual
flux measurements, along with flux measurements in random positions on
nearby galaxies and the same host galaxy.

The second candidate precursor is detected $\sim 700$\,days
prior to the SN explosion, and it is also based on two marginal
detections separated by about two weeks.
If real, the peak absolute $R$-band magnitude of this
precursor is about $-13$.

\begin{figure}
\centerline{\includegraphics[width=7.5cm]{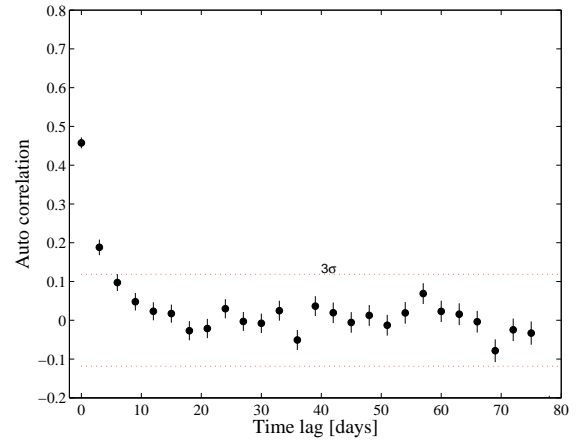}}
\caption{The discrete autocorrelation function (Edelson \& Krolik 1988) of the flux residuals
of PTF\,12cxj, taken before $t_{{\rm rise}}$.
In order to calculate the autocorrelation function we use time bins of 3\,days .
The dotted horizontal lines represent the lower and upper 3$\sigma$ bounds, estimated assuming $\sigma\approx1/\sqrt{N}$,
where $N$ is the number of measurements in each time bin.
\label{fig:PTF12cxj_ACFresid}}
\end{figure}

For all of the SNe in our sample, we also calculated the autocorrelation function
of the flux residuals prior to $t_{{\rm rise}}$.
The only SNe which exhibited significant (at the 3$\sigma$ level)
auto-correlation at lag one (i.e., corresponding to two successive measurements)
are PTF\,12cxj (4.7$\sigma$) and PTF\,10tel (3.2$\sigma$).
Figure~\ref{fig:PTF12cxj_ACFresid} presents the 
discrete autocorrelation function (Edelson \& Krolik 1988)
of all the flux residuals
of PTF\,12cxj,
measured before $t_{{\rm rise}}$.
The figure shows significant autocorrelation on time scales of a few days
to ten days.
This may indicate that the flux residuals are not pure
uncorrelated noise, but contain
a fraction of the progenitor light.
Moreover, it is possible that the progenitor is variable on time scales of a few days.
An alternative explanation to the variability
is that this signal is caused by the lunar synodic period
(i.e., the limiting magnitude is better during dark time).
Nevertheless, this means that the progenitor of PTF\,12cxj is
likely detected in several binned images,
and that its absolute $R$-band magnitude is about $-13$,
brighter even than the possible progenitor of SN\,2010jl/PTF\,10aaxf (Smith et al. 2011).
Given the detection of a signal in the autocorrelation,
we consider the two precursor events as real,
and their properties are listed in Table~\ref{tab:PrecProp}.

\section{Control time}
\label{sec:Control}

In order to calculate the rate of SN precursors,
we need to estimate the ``control time'' --- that is, for how long
each SN location was observed (prior to its explosion)
to a given limiting magnitude.
Table~\ref{tab:Table_ControlTime} lists,
for each SN, the time bin windows (of 15\,days)
prior to the SN explosion and the 5$\sigma$ sensitivity depth
at each window for bins with more than five measurements (second channel),
or the median 6$\sigma$ limiting magnitude at windows
with fewer than six measurements (first channel).
\begin{deluxetable}{lllll}
\tablecolumns{5}
\tablewidth{0pt}
\tablecaption{Control time}
\tablehead{
\colhead{Name}                &
\colhead{$\Delta{t}$}         &
\colhead{$R_{{\rm PTF}}$}       &
\colhead{Abs. $R_{{\rm PTF}}$}  &
\colhead{$N_{{\rm meas}}$}      \\
\colhead{}                    &
\colhead{(day)}                 &
\colhead{(mag)}                 &
\colhead{(mag)}                 &
\colhead{}            
}
\startdata
      PTF10aaxf&$   -228.6$& 21.4&$ -12.0$&    6\\
      PTF10aaxf&$   -203.8$& 20.9&$ -12.6$&    1\\
      PTF10aaxf&$   -192.3$& 20.0&$ -13.4$&    1\\
      PTF10aaxf&$   -179.8$& 20.7&$ -12.7$&    1\\
      PTF10aazn&$   -306.6$& 21.5&$ -12.8$&    8
\enddata
\tablecomments{Precursor search limiting magnitude in time windows.
$\Delta{t}$ is the mean time of the measurements
within each 15-day bin as measured
relative to $t_{{\rm rise}}$ of the SN (see Table~\ref{tab:Samp}).
$N_{{\rm meas}}$ is the number of data points in each 15-day bin.
For bins with $N_{{\rm meas}}<6$ we present the median
of all $6\sigma$ limiting magnitude in the bin.
We mark instances with $N_{{\rm meas}}=1$ (even if $N_{{\rm meas}}>1$).}
\label{tab:Table_ControlTime}
\end{deluxetable}

\begin{figure}
\centerline{\includegraphics[width=7.5cm]{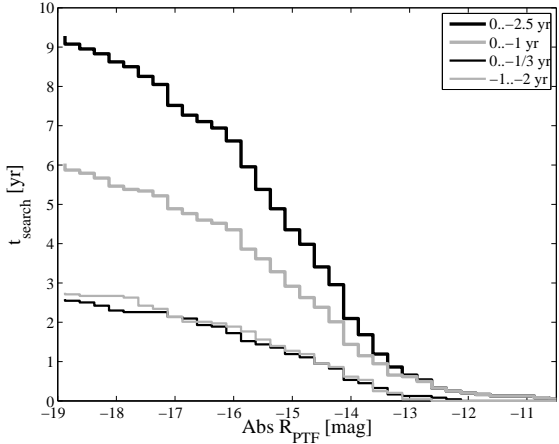}}
\caption{The cumulative precursor search control time in years as a function
of absolute magnitude.
For each time range prior to the SN explosion (see legend), this plot
shows the total number of years  we searched our entire sample for precursors
down to a given limiting magnitude (see text for details).
\label{fig:ControlTime}}
\end{figure}

To calculate the control time as a function
of absolute magnitude, we sum over all the SNe the number of
time bins (in Table~\ref{tab:Table_ControlTime})
in which the limiting absolute magnitude
is deeper than the absolute magnitude of interest,
and multiply by the bin size (i.e., 15\,days).
Figure~\ref{fig:ControlTime} displays
the sample cumulative control time as a function
of absolute $R$-band magnitude.
Specifically, this plot shows for several time
ranges (e.g., 0 to 2.5\,yr prior to the SN explosion; black line)
the total number of years, within a time range, in which we are able to detect
a precursor event brighter than a given absolute magnitude.

\begin{figure}
\centerline{\includegraphics[width=7.5cm]{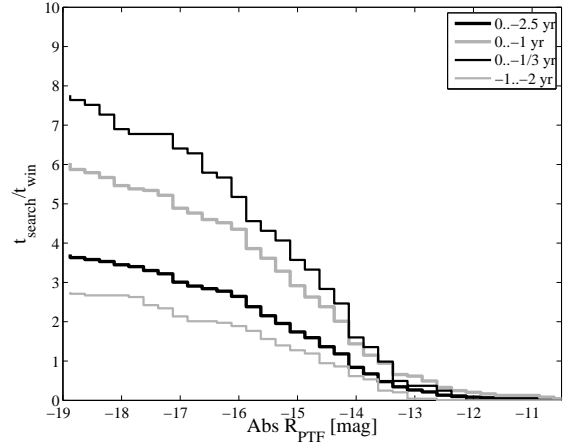}}
\caption{The cumulative precursor search control time fraction as a function
of absolute magnitude (Eq.~\ref{eq:fmr}).
For each time range prior to the SN explosion (see legend), this plot
shows the total number of SNe we observed for the equivalent
of the entire search window down to a given limiting magnitude.
\label{fig:ControlTimeFraction}}
\end{figure}

Figure~\ref{fig:ControlTimeFraction} shows
the sample cumulative control time fraction as a function
of absolute $R$-band magnitude.
This is defined as the total amount of time
(given in Fig.~\ref{fig:ControlTime})
divided by the size of the time window (listed in the legend).
This fraction represents the equivalent number of SNe observed
during the entire time window as a function
of absolute limiting magnitude.
The plot demonstrates that the highest efficiency is available for relatively
short periods prior to the SN explosion.
This is due to the fact that PTF switches many of the observed fields
typically every 3--4\,months, thereby probably explaining why many of 
the precursors we find are within 100\,days before the SN explosion.

\section{Precursor Rate}
\label{sec:rate}

Given the cumulative control time $t_{{\rm search}}(<M_R)$
over all SNe as a function of absolute magnitude ($M_R$), and
the cumulative control time fraction
\begin{equation}
f_{{\rm search}}(<M_R)= \frac{t_{{\rm search}}(<M_R)}{t_{{\rm win}}},
\label{eq:fmr}
\end{equation}
where $t_{{\rm win}}$ is the length of the search window considered (e.g.,
$1/3$\,yr prior to the explosion), we can calculate the {\it average}
(over all SNe) rate of precursors.
The mean precursor rate per SN,
in units of events per unit time, is
\begin{equation}
R_{{\rm prec}} = \frac{N_{{\rm prec}} }{(t_{{\rm search}}[<M_R])},
\label{eq:Rp}
\end{equation}
where $N_{{\rm prec}}$ is the total number of precursor events detected
in the sample within $t_{{\rm win}}$ (this can be more than one
precursor per SN).
The fraction of SNe~IIn that show at least one precursor event
within $t_{{\rm win}}$ is given by
\begin{equation}
f_{{\rm prec}} = \frac{N_{{\rm SN,prec}} }{(f_{{\rm search}}[<M_R])}.
\label{eq:fp}
\end{equation}
Here, $N_{{\rm SN,prec}}$ is the number of SNe that show at least one
precursor event within $t_{{\rm win}}$
(i.e., each SN with detected precursors
is counted once regardless of the number of precursor events).

Remarkably, we find that the fraction of SNe~IIn
that have precursors is
of order unity.  Assuming a homogeneous population, at the
one-sided 99\% confidence level, more than $52\%$ ($98\%$) of
SNe~IIn have at least one pre-explosion outburst
brighter than an
absolute magnitude of $-14$, and that takes place up to 1/3 (2.5)\,yr
prior to the SN explosion.  Furthermore, our results suggest that,
typically, SNe~IIn exhibit more than one precursor on a time scale of
1\,yr prior to the explosion.  Specifically, during this final year
prior to the explosion, the average rate of precursors brighter than absolute
magnitude $-14$,
is $\mathcal{R}$$=7.5_{-3.6,-5.5}^{+5.9,+12.0}$\,yr$^{-1}$
(1$\sigma$ and 2$\sigma$ errors).
We note that this estimate contains only Poisson errors,
and does~not include the prior that the number of events
cannot exceed one year divided by the typical duration
of each event.
Furthermore, even if we assume that the precursors of PTF\,10weh
and PTF\,12cxj are not real
(i.e., excluding events detected only in coadded images),
the derived rate is
$\mathcal{R}$$=3.8_{-2.4,-3.3}^{+4.9,+10.0}$\,yr$^{-1}$,
still likely above unity.

Our analysis also suggests that fainter precursors are even more common,
and that the cumulative luminosity function of precursors
is roughly $\mathcal{R}$$(>L_{{\rm peak}})\propto L_{{\rm peak}}^{-0.7\pm0.5}$,
where $\mathcal{R}$$(>L_{{\rm peak}})$
is the rate of precursors brighter than peak luminosity $L_{{\rm peak}}$.
We note that these
statistical results are in accord with individual
well-studied cases like SN\,2009ip, in which several precursors were
documented (e.g., Prieto et al. 2013; Margutti et al. 2013).

The derived precursor rate and fraction
of SNe with precursors, along with their 1$\sigma$ and 2$\sigma$
errors (Gehrels 1986), are listed in Table~\ref{tab:FreqPrec};
it suggests that the rate of precursors is larger than
unity. This may indicate that on average, each SN has more than one
precursor event in the few years prior to their explosion.
This is supported by cases like SN\,2009ip in which several
precursors were documented (e.g., Prieto et al. 2013).

\begin{deluxetable*}{llllllll}
\tablecolumns{8}
\tablewidth{0pt}
\tablecaption{Rate of Precursors}
\tablehead{
\colhead{$N_{{\rm SN,prec}}$}            &
\colhead{$N_{{\rm prec}}$}              &
\colhead{$M_R$}                 &
\colhead{$t_{{\rm win}}$}               &
\colhead{$t_{{\rm search}}(<M_R)$} &
\colhead{$f_{{\rm search}}(<M_R)$} &
\colhead{$\mathcal{R}$$_{{\rm prec}}(<M_R)$} &
\colhead{$f_{{\rm prec}}(<M_R)$}  \\
\colhead{}                  &
\colhead{}                  &
\colhead{(mag)}               &
\colhead{(yr)}                &
\colhead{(yr)}                &
\colhead{}                  &
\colhead{(yr$^{-1}$)}          &
\colhead{}
}
\startdata
 4 &  4 & $-14.0$ & 0.00..$-0.33$ &  0.53 &  1.60 & $  7.49_{ -3.59, -5.51}^{ +5.92,+11.96}$ & $  2.50_{ -1.20, -1.84}^{ +1.97, +3.99}$\\
 4 &  4 & $-14.0$ & 0.00..$-1.00$ &  1.44 &  1.44 & $  2.78_{ -1.33, -2.05}^{ +2.20, +4.44}$ & $  2.78_{ -1.33, -2.05}^{ +2.20, +4.44}$\\
 4 &  4 & $-14.0$ & 0.00..$-2.50$ &  2.09 &  0.84 & $  1.91_{ -0.91, -1.40}^{ +1.51, +3.05}$ & $  4.77_{ -2.29, -3.51}^{ +3.78, +7.62}$\\
 5 &  5 & $-11.8$ & 0.00..$-1.00$ &  0.16 &  0.16 & $ 30.44_{-13.15,-20.80}^{+20.59,+41.52}$ & $ 30.44_{-13.15,-20.80}^{+20.59,+41.52}$\\
 5 &  6 & $-11.8$ & 0.00..$-2.50$ &  0.16 &  0.07 & $ 36.52_{-14.49,-23.42}^{+21.82,+43.95}$ & $ 76.09_{-32.87,-52.00}^{+51.48,+103.80}$\\
\hline
 2 &  2 & $-14.0$ & 0.00..$-0.33$ &  0.53 &  1.60 & $3.75_{ -2.42, -3.32}^{ +4.94,+10.02}$   & $  1.25_{ -0.81, -1.16}^{ +1.65, +4.00}$\\
 2 &  2 & $-14.0$ & 0.00..$-1.00$ &  1.44 &  1.44 & $  1.39_{ -0.90, -1.23}^{ +1.84, +3.72}$ & $  1.39_{ -0.90, -1.29}^{ +1.84, +4.46}$
\enddata
\tablecomments{Average fraction of SNe~IIn
with precursors ($f_{{\rm prec}}[<M_R]$)
and rate of precursors ($\mathcal{R}$$_{{\rm prec}}[<M_R]$)
based on SNe in our sample.
The fractions and rates are listed for various time windows
prior to the SN explosion and different
absolute limiting magnitude ($M_R$).
The 2$\sigma$ and 3$\sigma$ limit refers to two-sided probabilities (Gehrels 1986).
The number of events listed below the horizontal line assumes that the
precursors detected in binned data (i.e., from PTF\,10weh and PTF\,12cxj)
are not real.
}
\label{tab:FreqPrec}
\end{deluxetable*}

A disadvantage of using the combined control time
of all the SNe is that it assumes that the precursor
properties are universal among all SNe~IIn.
Another caveat is that our search is mostly sensitive
 to events with a duration longer than about two weeks.
Shorter events can be detected but their control time
suffers from additional uncertainties.
In any case, our approach 
still gives us an estimate for the rate of precursors.

\section{CSM Mass Estimate}
\label{sec:CSM}

A possible scenario is that the precursors are associated
with mass-loss events.
Two scenarios can then be considered.
In the first scenario,
part of the CSM kinetic energy
is converted into luminosity. We therefore assume
that up to an efficiency factor $\epsilon$,
the kinetic energy of the CSM is equivalent
to its radiated energy.
In this case,
\begin{equation}
M_{{\rm CSM}} \approx \epsilon \frac{2L_{{\rm prec}}\delta{t}}{v^{2}}.
\label{eq:Mcsm}
\end{equation}
Here $L_{{\rm prec}}$ is the mean luminosity of the outburst,
$\delta{t}$ is its duration,
and $v$ is the velocity of the CSM.
We estimated the CSM velocity based on the width of the
wide component of the H$\alpha$ emission line
detected in the SN spectra (Table~\ref{tab:PrecProp}).

In an alternative scenario, the CSM and precursor can still be
related, but in an opposite way. Instead of transferring energy
from the CSM to the precursor, the CSM could be the result of the
precursor luminosity. This is naturally obtained if the CSM was
accelerated through a continuum-driven wind, such that
\begin{equation}
M_{{\rm CSM,cont}} \approx W \frac{L_{{\rm prec}}\delta{t}}{c_{{\rm s}}c}.
\label{eq:Mcsm_cont}
\end{equation}
Here $c_{{\rm s}}$ is the speed of sound at the base
of the optically thick wind ($\sim60$\,km\,s$^{-1}$),
$c$ is the speed of light, and $W\approx5$ is an empirically
derived constant (Shaviv 2000; 2001).
Although the flow of energy in the two scenarios is opposite, we 
note that the second scenario is equivalent to the first
with a universal velocity
and it does~not depend on the CSM velocity via the measured line widths.

\section{Correlation Between the Precursor and SN Properties}
\label{sec:Corr}

The properties of the candidate precursors and their
corresponding SN light curves are presented
in Table~\ref{tab:PrecProp}.
We also added to this table SN\,2009ip, for which similar
outbursts were detected and characterized.
If these SN light curves are powered by shock breakout
followed by interaction of the SN ejecta
with a dense CSM,
then we expect that some of the outburst and SN
properties will be correlated
(Ofek et al. 2010; 2014a; 2014b; Svirski et al. 2012).
If a substantial fraction of the SN radiated energy originates
from the interaction of the SN ejecta with the CSM,
then this material mass will determine the total (and peak)
radiated luminosity of the SN.
Moreover, the CSM
mass will determine the diffusion time scale
and therefore the SN rise time. 
Given this prediction, we search for such correlations.

\begin{figure}
\centerline{\includegraphics[width=7.5cm]{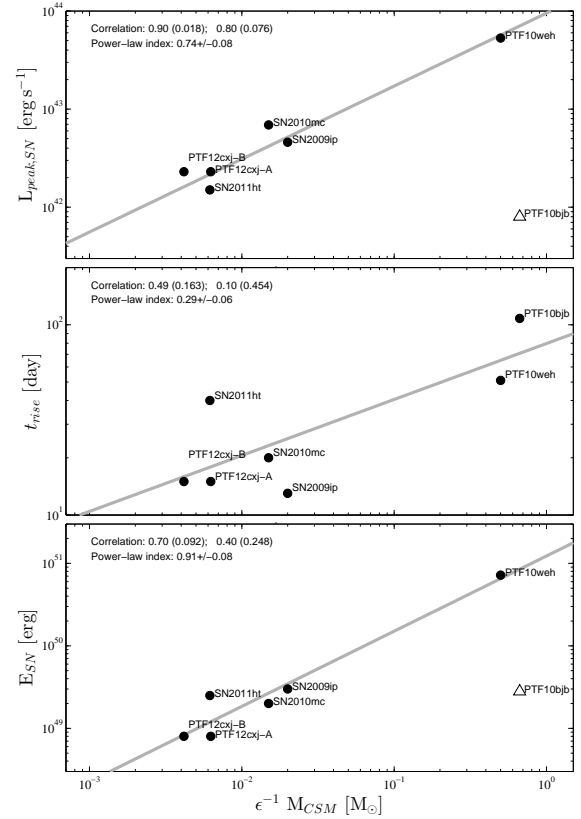}}
\caption{Correlations between the SN properties and
the ejected CSM mass estimated using Equation~\ref{eq:Mcsm}.
{\bf Top}: SN $R$-band peak luminosity vs. CSM mass.
{\bf Middle}: SN rise time vs. CSM mass.
{\bf bottom}: SN integrated $R$-band luminosity vs. CSM mass.
At the top of each panel, we also provide the Spearmann rank
correlation coefficient
and the probability of getting this correlation by chance.
This false-alarm probability was calculated using the bootstrap
technique (Efron 1982).
Two correlations are given: the first is for the entire sample,
while the second excludes PTF\,10weh.
The best-fit power law is shown with a gray line,
and the power-law index is provided below
the correlation coefficient.
The correlation and power law were calculated excluding
the PTF\,12cxj-B event. 
Excluding the PTF\,12cxj-A event or
combining the two points of PTF\,12cxj
into one (i.e., by adding the estimated CSM mass)
have a small impact on the results.
We note that the relative errors on each data point are estimated
to be on the order of tens of percent.
However, the use of the bootstrap technique gives us a robust estimate
for the false-alarm probability, regardless of the poorly known errors.
%
Furthermore, if we exclude PTF\,12cxj and PTF\,10weh (detected
via the second channel), then the sample is too small
to detect any correlations.
Therefore, a larger sample is required in order to verify these
correlations.
\label{fig:Mcsm_corr}}
\end{figure}

\begin{figure}
\centerline{\includegraphics[width=7.5cm]{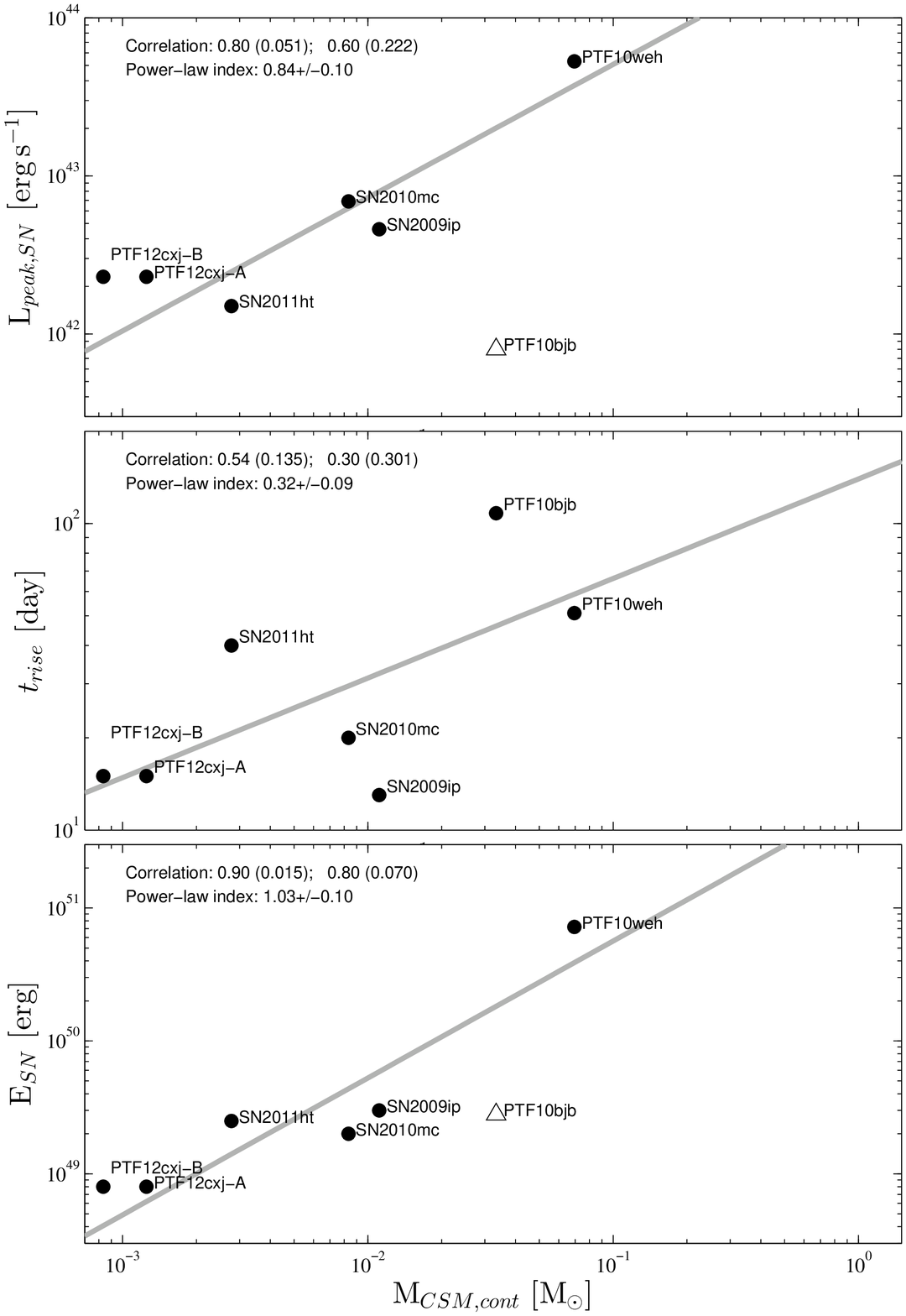}}
\caption{Like Figure~\ref{fig:Mcsm_corr}, but with the CSM mass estimated
based on  Equation~\ref{eq:Mcsm_cont}.
\label{fig:Mcsmcont_corr}}
\end{figure}

To estimate the
CSM mass, we used Equations~\ref{eq:Mcsm}
or \ref{eq:Mcsm_cont}.
We note that this approach assumes that $\epsilon$
is of the same order of magnitude for all the
precursors in our sample.
The three panels in Figure~\ref{fig:Mcsm_corr}
show (from top to bottom)
the precursor-ejected CSM mass
based on Equation~\ref{eq:Mcsm}, as a function of
the SN $R$-band peak luminosity, rise time, and integrated
$R$-band energy.
At the top of each panel, we also provide the
Spearmann rank correlation coefficient
and the probability of getting this correlation by chance.
Figure~\ref{fig:Mcsmcont_corr} shows the same, but for the CSM mass
estimated based on Equation~\ref{eq:Mcsm_cont}.

The clearest possible correlations we find are between
(i) Equation~\ref{eq:Mcsm} estimated CSM mass and the SN
peak luminosity;
and (ii) Equation~\ref{eq:Mcsm_cont} estimated CSM mass and
the total SN radiated energy.
The false-alarm probabilities for these correlations are lower than 2\%.
However, these correlations are based on a small sample
of objects whose properties were roughly measured.
Therefore, a confirmation of this correlation requires
a larger sample for which the properties of both SNe
and precursors are accurately measured.

At this time, our observations cannot distinguish
between scenarios in which the
radiation drives a vigorous wind
or an explosion drives a mass ejection.
Nevertheless, these correlations imply that at least the early optical light
curves of SNe~IIn are powered by shock breakout in an optically thick
CSM followed by interaction of the SN ejecta
with the CSM
(Ofek et al. 2010; Chevalier \& Irwin 2011).
The correlations we find between the SN and precursor
properties also suggest that precursors are indeed
accompanied by mass-loss events larger than
$\sim10^{-2}$\,M$_\odot$ (Figure~\ref{fig:Mcsm_corr}).
Finally, we caution that given the small size of the sample, the reality of these correlations
requires a verification based on a larger and better sample.

\section{Discussion}
\label{sec:Disc}

We present the first systematic search for precursors
prior to the explosion of Type IIn SNe.
Our search yeilds five SNe for which at least one precursor
is detected.
Based on the sample control time, we are able to calculate the
precursor rate.
Our observations suggest that the rate of precursors
brighter than an absolute magnitude of $-14$ in the last year
prior to the SN~IIn explosion is $\gtorder 1$\,yr$^{-1}$.
However, an important caveat is that
some of our putative SN explosions might not actually be 
terminal events in which the star ends its life. 
Indeed, for the case of SN\,2009ip, there have been
some suggestions that the latest detected outburst is
not the final SN explosion (e.g., Pastorello et al. 2013).
We note that it will be possible to test this hypothesis
using {\it HST} imaging by checking if the progenitor is still visible.

Next, we discuss some of the implications of this survey,
assuming that the SNe in our sample do indeed represent
the terminal explosion of their progenitors.
Observations of SN\,2010mc (Ofek et al. 2013b) suggest
that, given the short period between the precursor and the explosion
of SN\,2010mc, the two events are probably causally connected, in the
sense that the precursor is not random but instead deeply related to
the late stages of stellar evolution.
Following Ofek et al. (2013b),
the remarkable rate of these precursors can be used to relate their
nature to physical processes taking place during the final stages of
stellar evolution.
Naively, assuming that each outburst
releases $\sim10^{-2}$\,M$_{\odot}$
(see \S\ref{sec:CSM}; Ofek et al. 2013b, 2013c)
and that the mass of the progenitor
is on the order of 50\,M$_{\odot}$,
no more than $\sim5000$ outburst events can occur during the life
of the star.  For a progenitor lifetime of $\sim10^{7}$\,yr,
if the outbursts occur regularly during the star's life, then the
average time interval between outbursts would be less than
$\sim2000$\,yr. 

However, two observations make this scenario
unlikely. The first is that if every SN~IIn has 5000
precursors, then the observed rate in nearby galaxies would be
two orders of magnitude higher than the SN rate.
Such a high rate of precursors
(e.g., ``SN impostor''; Van-Dyk \& Matheson 2012)
is not seen, so it is likely that the
number of precursors per SN is much lower.
The second is that our results suggest that on order unity of
these progenitors have precursors within a year prior to explosion,
which means that the outbursts are much more likely to be physically
connected to a process occurring near the end of a star's life,
and probably not more than $\sim5000$\,yr ($1/2000$ of their lifetime)
prior to the SN explosion.
For massive stars this is related to processes that occur
after the beginning of carbon burning
(Woosley et al. 2002).
Furthermore, if each progenitor generates only a few outbursts
(rather than the strict upper limit of $\sim5000$),
then the physical process that is responsible for these
outbursts likely takes place after the beginning
of neon or oxygen burning
(a few months to a few years prior to the explosion).

Finally, we note that mass-loss events are likely not limited
to Type IIn SN progenitors; 
there is some evidence that other kinds of SNe
have precursors prior to the explosion.
For example, Corsi et al. (2013) reported on a possible
eruption of the Type Ic SN PTF\,11qcj about 2\,yr
prior to the SN explosion.
Gal-Yam et al. (2014) and Yaron et al. (in prep.)
show that spectra of some SNe obtained hours after the explosion
reveal narrow, high-excitation emission lines that likely
originate from a CSM that was ejected months to years
prior to the SN explosion.
The main difference is probably that in SNe~IIn,
the ejected mass is larger than in other classes of events.

\acknowledgments

E.O.O. thanks Orly Gnat and Ehud Nakar for many discussions.
This paper is based on observations obtained with the
Samuel Oschin Telescope as part of the Palomar Transient Factory
project, a scientific collaboration between the
California Institute of Technology,
Columbia University,
Las Cumbres Observatory,
the Lawrence Berkeley National Laboratory,
the National Energy Research Scientific Computing Center,
the University of Oxford, and the Weizmann Institute of Science.
Some of the data presented herein were obtained at the W. M. Keck
Observatory, which is operated as a scientific partnership among the
California Institute of Technology, the University of California,
and NASA; the Observatory was made possible by the generous
financial support of the W. M. Keck Foundation.  We are grateful for
excellent staff assistance at Palomar, Lick, and Keck Observatories.
E.O.O. is incumbent of
the Arye Dissentshik career development chair and
is grateful for support by
a grant from the Israeli Ministry of Science and
the I-CORE Program of the Planning
and Budgeting Committee and The Israel Science Foundation (grant No 1829/12).
A.V.F.'s group at UC Berkeley has received generous financial
assistance from Gary and Cynthia Bengier, the Christopher R. Redlich
Fund, the Richard and Rhoda Goldman Fund, the TABASGO Foundation,
and NSF grant AST-1211916.


\begin{thebibliography}{}


\bibitem[Arcavi et al.(2011)]{2011ApJ...742L..18A} Arcavi, I., Gal-Yam, A., 
Yaron, O., et al.\ 2011, \apjl, 742, L18 

\bibitem[Arnaud(1996)]{1996ASPC..101...17A} Arnaud, K.~A.\ 1996, 
Astronomical Data Analysis Software and Systems V, 101, 17 

\bibitem[Arnett 
\& Meakin(2011)]{2011ApJ...741...33A} Arnett, W.~D., \& Meakin, C.\ 2011, \apj, 741, 33 

\bibitem[Balberg 
\& Loeb(2011)]{2011MNRAS.414.1715B} Balberg, S., \& Loeb, A.\ 2011, \mnras, 414, 1715 

\bibitem[Benetti et al.(2010)]{2010CBET.2536....1B} Benetti, S., Bufano, 
F., Vinko, J., et al.\ 2010, Central Bureau Electronic Telegrams, 2536, 1 

\bibitem[Boles et al.(2011)]{2011CBET.2851....1B} Boles, T., Pastorello, 
A., Stanishev, V., et al.\ 2011, Central Bureau Electronic Telegrams, 2851, 
1

\bibitem[Brown et al.(2009)]{2009AJ....137.4517B} Brown, P.~J., Holland, 
S.~T., Immler, S., et al.\ 2009, \aj, 137, 4517 

\bibitem[Burrows et al.(2005)]{2005SSRv..120..165B} Burrows, D.~N., Hill, 
J.~E., Nousek, J.~A., et al.\ 2005, \ssr, 120, 165 

\bibitem[Cardelli et al.(1989)]{1989ApJ...345..245C} Cardelli, J.~A., 
Clayton, G.~C., \& Mathis, J.~S.\ 1989, \apj, 345, 245 

\bibitem[Challis et al.(2010)]{2010CBET.2243....1C} Challis, P., Foley, 
R.~J., \& Berlind, P.\ 2010, Central Bureau Electronic Telegrams, 2243, 1 


\bibitem[Chandra et al.(2012)]{2012ApJ...750L...2C} Chandra, P., Chevalier, 
R.~A., Irwin, C.~M., et al.\ 2012, \apjl, 750, L2 

\bibitem[Chevalier(1982)]{1982ApJ...259..302C} Chevalier, R.~A.\ 1982, 
\apj, 259, 302 

\bibitem[Chevalier(1998)]{1998ApJ...499..810C} Chevalier, R.~A.\ 1998, 
\apj, 499, 810 

\bibitem[Chevalier(2012)]{2012ApJ...752L...2C} Chevalier, R.~A.\ 2012, 
\apjl, 752, L2 

\bibitem[Chevalier(2013)]{2013arXiv1304.5500C} Chevalier, R.~A.\ 2013, 
arXiv:1304.5500 

\bibitem[Chevalier 
\& Fransson(1994)]{1994ApJ...420..268C} Chevalier, R.~A., \& Fransson, C.\ 1994, \apj, 420, 268 

\bibitem[Chevalier 
\& Irwin(2011)]{2011ApJ...729L...6C} Chevalier, R.~A., \& Irwin, C.~M.\ 2011, \apjl, 729, L6 

\bibitem[Chevalier 
\& Irwin(2012)]{2012arXiv1201.5581C} Chevalier, R.~A., \& Irwin, C.~M.\ 2012, arXiv:1201.5581 

\bibitem[Chugai et al.(2003)]{2003astro.ph..9226C} Chugai, N.~N., Cumming, 
R.~J., Blinnikov, S.~I., et al.\ 2003, arXiv:astro-ph/0309226 

\bibitem[Chugai 
\& Danziger(1994)]{1994MNRAS.268..173C} Chugai, N.~N., \& Danziger, I.~J.\ 1994, MNRAS, 268, 173 

\bibitem[Ciabattari et al.(2011)]{2011CBET.2830....1C} Ciabattari, F., 
Mazzoni, E., Koff, R.~A., et al.\ 2011, Central Bureau Electronic 
Telegrams, 2830, 1 

\bibitem[Corsi et al.(2013)]{2013arXiv1307.2366C} Corsi, A., Ofek, E.~O., 
Gal-Yam, A., et al.\ 2013, arXiv:1307.2366 

\bibitem[Dessart et al.(2009)]{2009MNRAS.394...21D} Dessart, L., Hillier, 
D.~J., Gezari, S., Basa, S., \& Matheson, T.\ 2009, \mnras, 394, 21 

\bibitem[Dickey \& Lockman(1990)]{1990ARA&A..28..215D} Dickey, J.~M., \& Lockman, F.~J.\ 1990, \araa, 28, 215 

\bibitem[Dintinjana et al.(2011)]{2011CBET.2906....1D} Dintinjana, B., 
Mikuz, H., Skvarc, J., et al.\ 2011, Central Bureau Electronic Telegrams, 
2906, 1 

\bibitem[Dopita et al.(1984)]{1984ApJ...287L..69D} Dopita, M.~A., Cohen, 
M., Schwartz, R.~D., \& Evans, R.\ 1984, ApJL, 287, L69 

\bibitem[Draine 
\& McKee(1993)]{1993ARA&A..31..373D} Draine, B.~T., \& McKee, C.~F.\ 1993, \araa, 31, 373 

\bibitem[Drake et al.(2009)]{2009ApJ...696..870D} Drake, A.~J., Djorgovski, 
S.~G., Mahabal, A., et al.\ 2009, \apj, 696, 870 

\bibitem[Drake et al.(2010)]{2010ATel.2897....1D} Drake, A.~J., Prieto, 
J.~L., Djorgovski, S.~G., et al.\ 2010, The Astronomer's Telegram, 2897, 1 

\bibitem[Duszanowicz(2010)]{2010CBET.2241....1D} Duszanowicz, G.\ 2010, 
Central Bureau Electronic Telegrams, 2241, 1 

\bibitem[Edelson 
\& Krolik(1988)]{1988ApJ...333..646E} Edelson, R.~A., \& Krolik, J.~H.\ 1988, \apj, 333, 646 

\bibitem[Efron(1982)]{Efron1982} Efron, B., 1982, The Jackknife, the Bootstrap and Other Resampling Plans, (The Society for Industrial and Applied Mathematics)

\bibitem[ET(1993)]{ET1993} Efron, B., Tibshirani, R. J., 1993,
An Introduction to the Bootstrap, Monographs on Statistics and Applied Probability 57 (Chapman \& Hall/CRC) ISBN 0-412-04231-2

\bibitem[Falk \& Arnett(1973)]{1973ApJ...180L..65F} Falk, S.~W., \& Arnett, W.~D.\ 1973, \apjl, 180, L65 

\bibitem[Falk 
\& Arnett(1977)]{1977ApJS...33..515F} Falk, S.~W., \& Arnett, W.~D.\ 1977, \apjs, 33, 515 

\bibitem[Filippenko(1997)]{1997ARA&A..35..309F} Filippenko, A.~V.\ 1997, ARA\&A, 35, 309 

\bibitem[Filippenko et al.(2001)]{2001ASPC..246..121F} Filippenko, A.~V., 
Li, W.~D., Treffers, R.~R., 
\& Modjaz, M.\ 2001, in Small-Telescope Astronomy on Global Scales,
 W. P. Chen, C. Lemme, \& B. Paczy\'{n}ski (San Francisco: ASP,
   Conf. Ser. Vol. 246), 121

\bibitem[Foley et al.(2011)]{2011ApJ...732...32F} Foley, R.~J., Berger, E., 
Fox, O., et al.\ 2011, \apj, 732, 32 

\bibitem[Foley et al.(2007)]{2007ApJ...657L.105F} Foley, R.~J., Smith, N., 
Ganeshalingam, M., et al.\ 2007, \apjl, 657, L105 

\bibitem[Fraser et al.(2013)]{2013arXiv1309.4695F} Fraser, M., Magee, M., 
Kotak, R., et al.\ 2013, arXiv:1309.4695 

\bibitem[Gal-Yam 
\& Leonard(2009)]{2009Natur.458..865G} Gal-Yam, A., \& Leonard, D.~C.\ 2009, \nat, 458, 865 

\bibitem[Gal-Yam et al.(2007)]{2007ApJ...656..372G} Gal-Yam, A., Leonard, 
D.~C., Fox, D.~B., et al.\ 2007, \apj, 656, 372 

\bibitem[Gal-Yam(2012)]{2012Sci...337..927G} Gal-Yam, A.\ 2012, Science, 
337, 927 

\bibitem[Gal-Yam et al.(2013)]{GY2013} Gal-Yam, A., et al.\ 2014, submitted

\bibitem[Gehrels(1986)]{1986ApJ...303..336G} Gehrels, N.\ 1986, \apj, 303, 
336 

\bibitem[Gehrels et al.(2004)]{2004ApJ...611.1005G} Gehrels, N., 
Chincarini, G., Giommi, P., et al.\ 2004, \apj, 611, 1005 

\bibitem[Ghavamian et al.(2013)]{2013arXiv1305.6617G} Ghavamian, P., 
Schwartz, S.~J., Mitchell, J., Masters, A., 
\& Laming, J.~M.\ 2013, arXiv:1305.6617 

\bibitem[Ginzburg \& Balberg(2012)]{2012arXiv1205.3455G} Ginzburg, S., \& Balberg, S.\ 2012, arXiv:1205.3455 

\bibitem[Gnat 
\& Sternberg(2009)]{2009ApJ...693.1514G} Gnat, O., \& Sternberg, A.\ 2009, \apj, 693, 1514 

\bibitem[Harrison et al.(2013)]{2013ApJ...770..103H} Harrison, F.~A., 
Craig, W.~W., Christensen, F.~E., et al.\ 2013, \apj, 770, 103 


\bibitem[Horesh et al.(2012)]{2012XXX} Horesh, A., et al. 2012, submitted to ApJ

\bibitem[Itoh(1978)]{1978PASJ...30..489I} Itoh, H.\ 1978, \pasj, 30, 489 


\bibitem[Jin et al.(2012)]{2012CBET.3044....1J} Jin, Z., Gao, X., 
Brimacombe, J., Luppi, F., 
\& Buzzi, L.\ 2012, Central Bureau Electronic Telegrams, 3044, 1 

\bibitem[Kasen 
\& Bildsten(2010)]{2010ApJ...717..245K} Kasen, D., \& Bildsten, L.\ 2010, \apj, 717, 245 

\bibitem[Katz et al.(2011)]{2011arXiv1106.1898K} Katz, B., Sapir, N., 
\& Waxman, E.\ 2011, arXiv:1106.1898 

\bibitem[Kiewe et al.(2012)]{2012ApJ...744...10K} Kiewe, M., Gal-Yam, A., 
Arcavi, I., et al.\ 2012, ApJ, 744, 10 

\bibitem[Lang(1999)]{1999acfp.book.....L} Lang, K.~R.\ 1999, Astrophysical 
Formulae (New York: Springer)

\bibitem[Law et al.(2009)]{2009PASP..121.1395L} Law, N.~M., et al.\ 2009, PASP, 121, 1395 

\bibitem[Li et al.(2000)]{2000AIPC..522..103L} Li, W.~D., Filippenko, 
A.~V., Treffers, R.~R., et al.\ 2000, in Cosmic Explosions, ed. S. S. Holt 
\& W. W. Zhang (New York: AIP), 103

\bibitem[Mahabal et al.(2011)]{2011CBET.2941....1M} Mahabal, A., Drake, 
A.~J., Djorgovski, S.~G., et al.\ 2011, Central Bureau Electronic 
Telegrams, 2941, 1 

\bibitem[Margutti et al.(2014)]{2014ApJ...780...21M} Margutti, R., 
Milisavljevic, D., Soderberg, A.~M., et al.\ 2014, \apj, 780, 21 

\bibitem[Mason et al.(2011)]{2011CBET.2712....1M} Mason, M., Cenko, S.~B., 
Li, W., et al.\ 2011, Central Bureau Electronic Telegrams, 2712, 1 

\bibitem[Matzner \& McKee(1999)]{1999ApJ...510..379M} Matzner, C.~D., \& McKee, C.~F.\ 1999, ApJ, 510, 379 

\bibitem[Mauerhan et al.(2013a)]{2013MNRAS.430.1801M} Mauerhan, J.~C., 
Smith, N., Filippenko, A.~V., et al.\ 2013a, \mnras, 430, 1801 

\bibitem[Mauerhan et al.(2013b)]{2013MNRAS.431.2599M} Mauerhan, J.~C., 
Smith, N., Silverman, J.~M., et al.\ 2013b, \mnras, 431, 2599 

\bibitem[Moriya et al.(2013)]{2013MNRAS.435.1520M} Moriya, T.~J., Maeda, 
K., Taddia, F., et al.\ 2013, \mnras, 435, 1520 

\bibitem[Moriya 
\& Tominaga(2012)]{2012ApJ...747..118M} Moriya, T.~J., \& Tominaga, N.\ 2012, \apj, 747, 118 

\bibitem[Morrison \& McCammon(1983)]{1983ApJ...270..119M} Morrison, R., \& McCammon, D.\ 1983, \apj, 270, 119 

\bibitem[Murase et al.(2011)]{2011PhRvD..84d3003M} Murase, K., Thompson, 
T.~A., Lacki, B.~C., \& Beacom, J.~F.\ 2011, \prd, 84, 043003 

\bibitem[Nakar 
\& Sari(2010)]{2010ApJ...725..904N} Nakar, E., \& Sari, R.\ 2010, ApJ, 725, 904 
\bibitem[Newton 
\& Puckett(2010)]{2010CBET.2532....1N} Newton, J., \& Puckett, T.\ 2010, Central Bureau Electronic Telegrams, 2532, 1 

\bibitem[Nomoto 
\& Sugimoto(1972)]{1972PThPh..48...46N} Nomoto, K., \& Sugimoto, D.\ 1972, Progress of Theoretical Physics, 48, 46 

\bibitem[Nugent et al.(2011)]{2011Natur.480..344N} Nugent, P.~E., Sullivan, 
M., Cenko, S.~B., et al.\ 2011, \nat, 480, 344 



\bibitem[Ofek(2012)]{2012CBET.3313....1O} Ofek, O.\ 2012, Central Bureau 
Electronic Telegrams, 3313, 1 

\bibitem[Ofek et al.(2007)]{2007ApJ...659L..13O} Ofek, E.~O., et al.\ 2007, 
ApJ, 659, L13 

\bibitem[Ofek et al.(2013a)]{2013ApJ...763...42O} Ofek, E.~O., Fox, D., 
Cenko, S.~B., et al.\ 2013a, \apj, 763, 42 

\bibitem[Ofek et al.(2012a)]{2012PASP..124...62O} Ofek, E.~O., Laher, R., 
Law, N., et al.\ 2012a, \pasp, 124, 62 

\bibitem[Ofek et al.(2012b)]{2012PASP..124..854O} Ofek, E.~O., Laher, R., 
Surace, J., et al.\ 2012b, \pasp, 124, 854 

\bibitem[Ofek et al.(2013c)]{2013ApJ...768...47O} Ofek, E.~O., Lin, L., 
Kouveliotou, C., et al.\ 2013c, \apj, 768, 47 

\bibitem[Ofek et al.(2010)]{2010ApJ...724.1396O} Ofek, E.~O., Rabinak, I., 
Neill, J.~D., et al.\ 2010, \apj, 724, 1396 

\bibitem[Ofek et al.(2013b)]{2013Natur.494...65O} Ofek, E.~O., Sullivan, M., 
Cenko, S.~B., et al.\ 2013b, \nat, 494, 65 

\bibitem[Ofek et al.(2014)]{2014ApJ...781...42O} Ofek, E.~O., Zoglauer, A., 
Boggs, S.~E., et al.\ 2014, \apj, 781, 42 

\bibitem[Pastorello et al.(2013)]{2013ApJ...767....1P} Pastorello, A., 
Cappellaro, E., Inserra, C., et al.\ 2013, \apj, 767, 1 

\bibitem[Pastorello et al.(2008)]{2008MNRAS.389..113P} Pastorello, A., 
Mattila, S., Zampieri, L., et al.\ 2008, \mnras, 389, 113 

\bibitem[Pastorello et al.(2007)]{2007Natur.447..829P} Pastorello, A., 
Smartt, S.~J., Mattila, S., et al.\ 2007, \nat, 447, 829 

\bibitem[Pastorello et al.(2011)]{2011CBET.2851....2P} Pastorello, A., 
Stanishev, V., Smartt, S.~J., Fraser, M., 
\& Lindborg, M.\ 2011, Central Bureau Electronic Telegrams, 2851, 2 

\bibitem[Patat et 
al.(2011)]{2011A&A...527L...6P} Patat, F., Taubenberger, S., Benetti, S., Pastorello, A., \& Harutyunyan, A.\ 2011, \aap, 527, L6 

\bibitem[Poole et al.(2008)]{2008MNRAS.383..627P} Poole, T.~S., Breeveld, 
A.~A., Page, M.~J., et al.\ 2008, \mnras, 383, 627 

\bibitem[Prieto et al.(2013)]{2013ApJ...763L..27P} Prieto, J.~L., 
Brimacombe, J., Drake, A.~J., \& Howerton, S.\ 2013, \apjl, 763, L27 

\bibitem[Prieto et al.(2011)]{2011CBET.2903....1P} Prieto, J.~L., McMillan, 
R., Bakos, G., 
\& Grennan, D.\ 2011, Central Bureau Electronic Telegrams, 2903, 1 

\bibitem[Quataert 
\& Shiode(2012)]{2012MNRAS.423L..92Q} Quataert, E., \& Shiode, J.\ 2012, MNRAS, 423, L92 


\bibitem[Quimby et al.(2011)]{2011Natur.474..487Q} Quimby, R.~M., Kulkarni, 
S.~R., Kasliwal, M.~M., et al.\ 2011, \nat, 474, 487 

\bibitem[Rakavy et al.(1967)]{1967ApJ...150..131R} Rakavy, G., Shaviv, G., 
\& Zinamon, Z.\ 1967, \apj, 150, 131

\bibitem[Rau et al.(2009)]{2009PASP..121.1334R} Rau, A., et al.\ 2009, PASP, 121, 1334 

\bibitem[Rest et al.(2011)]{2011ApJ...729...88R} Rest, A., Foley, R.~J., 
Gezari, S., et al.\ 2011, \apj, 729, 88 

\bibitem[Rich(2010)]{2010CBET.2530....1R} Rich, D.\ 2010, Central Bureau 
Electronic Telegrams, 2530, 1 

\bibitem[Roming et al.(2005)]{2005SSRv..120...95R} Roming, P.~W.~A., 
Kennedy, T.~E., Mason, K.~O., et al.\ 2005, \ssr, 120, 95 

\bibitem[Roming et al.(2012)]{2012arXiv1202.4840R} Roming, P.~W.~A., 
Pritchard, T.~A., Prieto, J.~L., et al.\ 2012, arXiv:1202.4840 

\bibitem[Schafer(1991)]{1991xxrs.book.....S} Schafer, R.~A.\ 1991, XSPEC, 
an X-ray Spectral Fitting Package: Version 2 of the User's Guide (Paris, 
France: European Space Agency)

\bibitem[Schlegel et al.(1998)]{1998ApJ...500..525S} Schlegel, D.~J., Finkbeiner, D.~P., \& Davis, M.\ 1998, ApJ, 500, 525 

\bibitem[Schlegel(1990)]{1990MNRAS.244..269S} Schlegel, E.~M.\ 1990, 
\mnras, 244, 269 

\bibitem[Shaviv(2000)]{2000ApJ...532L.137S} Shaviv, N.~J.\ 2000, \apjl, 
532, L137 

\bibitem[Shaviv(2001)]{2001MNRAS.326..126S} Shaviv, N.~J.\ 2001, \mnras, 
326, 126 

\bibitem[Shiode \& Quataert(2013)]{2013arXiv1308.5978S} Shiode, J.~H., \& Quataert, E.\ 2013, arXiv:1308.5978 

\bibitem[Silverman et al.(2010)]{2010CBET.2538....1S} Silverman, J.~M., 
Filippenko, A.~V., 
\& Foley, R.~J.\ 2010, Central Bureau Electronic Telegrams, 2538, 1 

\bibitem[Slysh(1990)]{1990SvAL...16..339S} Slysh, V.~I.\ 1990, Soviet 
Astronomy Letters, 16, 339 

\bibitem[Smith et al.(2008)]{2008ApJ...686..467S} Smith, N., Chornock, R., 
Li, W., et al.\ 2008, ApJ, 686, 467 

\bibitem[Smith et al.(2007)]{2007ApJ...666.1116S} Smith, N., Li, W., Foley, 
R.~J., et al.\ 2007, ApJ, 666, 1116 

\bibitem[Smith et al.(2011)]{2011ApJ...732...63S} Smith, N., Li, W., 
Miller, A.~A., et al.\ 2011, \apj, 732, 63 

\bibitem[Smith et al.(2010)]{2010AJ....139.1451S} Smith, N., Miller, A., 
Li, W., et al.\ 2010, AJ, 139, 1451 

\bibitem[Smith et al.(2009)]{2009ApJ...695.1334S} Smith, N., Silverman, 
J.~M., Chornock, R., et al.\ 2009, ApJ, 695, 1334 

\bibitem[Smith et al.(2012)]{2012AJ....143...17S} Smith, N., Silverman, 
J.~M., Filippenko, A.~V., et al.\ 2012, \aj, 143, 17 

\bibitem[Soker 
\& Kashi(2013)]{2013ApJ...764L...6S} Soker, N., \& Kashi, A.\ 2013, \apjl, 764, L6 

\bibitem[Stoll et al.(2011)]{2011ApJ...730...34S} Stoll, R., Prieto, J.~L., 
Stanek, K.~Z., et al.\ 2011, \apj, 730, 34 

\bibitem[Svirski et al.(2012)]{2012arXiv1202.3437S} Svirski, G., Nakar, E., 
\& Sari, R.\ 2012, arXiv:1202.3437 

\bibitem[Szczygie{\l} et al.(2012)]{2012ApJ...747...23S} Szczygie{\l}, 
D.~M., Gerke, J.~R., Kochanek, C.~S., \& Stanek, K.~Z.\ 2012, \apj, 747, 23 

\bibitem[Valenti et al.(2011)]{2011CBET.2906....2V} Valenti, S., 
Pastorello, A., Benetti, S., et al.\ 2011, Central Bureau Electronic 
Telegrams, 2906, 2 

\bibitem[Van Dyk 
\& Matheson(2012)]{2012ASSL..384..249V} Van Dyk, S.~D., \& Matheson, T.\ 2012, Astrophysics and Space Science Library, 384, 249 

\bibitem[Waldman(2008)]{2008ApJ...685.1103W} Waldman, R.\ 2008, \apj, 685, 
1103 

\bibitem[Waxman 
\& Shvarts(1993)]{1993PhFl....5.1035W} Waxman, E., \& Shvarts, D.\ 1993, Physics of Fluids, 5, 1035 

\bibitem[Weaver(1976)]{1976ApJS...32..233W} Weaver, T.~A.\ 1976, ApJS, 32, 233

\bibitem[Weiler et al.(1991)]{1991ApJ...380..161W} Weiler, K.~W., Van Dyk, 
S.~D., Discenna, J.~L., Panagia, N., \& Sramek, R.~A.\ 1991, \apj, 380, 161 

\bibitem[Woosley et al.(2007)]{2007Natur.450..390W} Woosley, S.~E., 
Blinnikov, S., \& Heger, A.\ 2007, Nature, 450, 390 

\bibitem[Woosley et al.(2002)]{2002RvMP...74.1015W} Woosley, S.~E., Heger, 
A., \& Weaver, T.~A.\ 2002, Reviews of Modern Physics, 74, 1015 

\bibitem[Yamanaka et al.(2010)]{2010CBET.2539....1Y} Yamanaka, M., 
Okushima, T., Arai, A., Sasada, M., 
\& Sato, H.\ 2010, Central Bureau Electronic Telegrams, 2539, 1 

\bibitem[Yaron 
\& Gal-Yam(2012)]{2012arXiv1204.1891Y} Yaron, O., \& Gal-Yam, A.\ 2012, arXiv:1204.1891 

\bibitem[Zhang et al.(2012)]{2012AJ....144..131Z} Zhang, T., Wang, X., Wu, 
C., et al.\ 2012, \aj, 144, 131 


\end{thebibliography}
\end{document}